\documentclass[preprint,12pt]{elsarticle}

\usepackage{amsmath}
\usepackage{amssymb}
\usepackage{multirow}%
\usepackage{amsthm}%
\usepackage{mathrsfs}%
\usepackage{algorithmicx}
\usepackage{amssymb}
\usepackage{lipsum}

\journal{Physics Letters A}
\begin{document}
\begin{frontmatter}
\title{Predicting positon solutions of a family of Nonlinear Schr\"{o}dinger equations through Deep Learning algorithm}
\author[1]{K. Thulasidharan}
\author[2]{N. Vishnu Priya}
\author[1]{S. Monisha}
\author[1]{M. Senthilvelan}
\affiliation[1]{organization={Department of Nonlinear Dynamics, Bharathidasan University}, 
            city={Tiruchirappalli},
            postcode={620 024}, 
            state={ Tamil Nadu},
            country={India}}
\affiliation[2]{organization={Department of Mathematics, Indian Institute of Science},
           city={Bengaluru},
            postcode={560 012},
           state={Karnataka},
            country={India}}
\begin{abstract}
We consider a hierarchy of nonlinear Schr{\"o}dinger equations (NLSEs) and forecast the evolution of positon solutions using a deep learning approach called Physics Informed Neural Networks (PINN). Notably, the PINN algorithm accurately predicts positon solutions not only in the standard NLSE but also in other higher order versions, including cubic, quartic and quintic NLSEs. The PINN approach also effectively handles two coupled NLSEs and two coupled Hirota equations. In addition to the above, we report exact second-order positon solutions of the sextic NLSE and coupled generalized NLSE. These solutions are not available in the existing literature and we construct them through generalized Darboux transformation method. Further, we utilize PINNs to forecast their behaviour as well. To validate PINN's accuracy, we compare the predicted solutions with exact solutions obtained from analytical methods. The results show high fidelity and low mean squared error in the predictions generated by our PINN model.
\end{abstract}


\end{frontmatter}

\section{Introduction}  
\par Study of localized wave solutions in nonlinear integrable systems is flourishing across various disciplines, including nonlinear optics,  fluid mechanics, astrophysics,  Bose-Einstein condensates, condensed matter physics, plasma physics and  oceanography \cite{soliton,chen, Ablowitz,agrawal}. Solitons are prominent solutions with diverse applications in these fields. Traditionally, solitons were constructed using negative eigenvalues in Lax pair equations via the Darboux transformation method \cite{Matveevbook}. Positon solutions refer to a specific type of solitary wave solutions characterized by positive eigenvalues, often seen as a counterpart to solitons.  As far as the  NLS-type equations are concerned, positons are also recognized as degenerate soliton solutions due to the degeneracy of complex spectral parameter, which travels with equal amplitude for larger values of time. Positon solution was introduced by Matveev by leveraging  generalized Darboux transformation (gDT) method for the Korteweg-de Vries (KdV) equation and considering positive eigenvalues in the Lax pair equations \cite{M1,M2,M3}. Positon solutions of the KdV equation are singular solutions that exhibit soliton-like behaviour in long range \cite{kdv}. Julia et al. overcame the singular nature of positon solutions by introducing a class of non-singular positon solutions known as smooth positons.  Notably, these smooth positons can model the tidal bore phenomenon in rivers \cite{tidal}. Subsequently, positon solutions have been formulated for various nonlinear evolutionary equations, encompassing NLSE, Hirota equation, complex modified KdV equation, generalized NLSE (Lakshmanan-Porsezian-Daniel equation), fifth order NLSE and so on \cite{Hirota2, w1, w2, fnls, cmkdv,  A1, kundu}. Further, breather positons are also constructed on plane wave background for several nonlinear equations \cite{w1,w2,bp-nls-mb,bp-KE}.   The aforementioned solutions move as a compound for a small time-period and travel separately in the larger time scale \cite{Hirota2}. Furthermore, positon solutions have been developed for specific two-component systems, such as the coupled NLSE (Manakov model) \cite{cnls} and coupled Hirota equation \cite{nld-paper}.

\par Recently, Machine Learning (ML) algorithms have been used to predict and analyze several phenomena, say for example, chaos \cite{ml-chaos}, extreme events \cite{meiy-ml2}, unstable periodic orbits \cite{ml-unstable} and chimera states \cite{sarika-chimera} in the field of nonlinear dynamics. The vast amount of data in nonlinear dynamics has motivated researchers to employ various artificial neural network architectures for predicting  nonlinear wave solutions of nonlinear partial differential equations (PDEs). In this direction, Raissi et~al. \cite{raissi-pinn} pioneered Physics-Informed Neural Networks (PINNs), a powerful deep learning method for tackling a wide range of PDEs. PINNs uniquely blend deep learning with physics principles for greater accuracy in scientific and engineering simulations. Crucially, PINNs are unsupervised, eliminating the need for labeled data from simulations or experiments.  PINNs recast PDE solutions as loss function optimization problems  \cite{pinn-review}. PINNs and their generalized forms have found application in diverse scientific fields, including fluid dynamics, nano-optics, metamaterials, and even modeling the spread of diseases. \cite{pinn-review,goswami,nano-optics,mol-dynamics,heat-transfer}. Researchers have applied PINNs to predict data-driven solutions like solitons, breathers, and rogue waves in nonlinear integrable equations \cite{pinn-kdv,meiy-ml3,dnls-pinn, hirota-pinn, 3cnls, tonls, zhenya1}. This study enhances the capabilities of PINN by investigating positon solutions across a hierarchy of NLSEs. We have successfully predicted second-order positons in NLSE families with cubic, quartic, quintic, and sextic nonlinearities, as well as in coupled NLSE, coupled Hirota equations and coupled generalized NLSE. In addition to the above, we report second-order positon solutions for two nonlinear PDEs, namely (i) sixth-order NLSE and (ii) coupled generalized NLSE. We derive the positon solutions using GDT method and these solutions are reported first time in the literature for these two equations. By incorporating these new solutions as initial data within the PINN algorithm, we also predict their behaviour with greater accuracy. Our findings demonstrate the neural network's ability to approximate positon solutions excellently with minimal error throughout the entire NLSE family. We mention here that the present work is the first instance of applying deep learning approach to predict positon solutions within this specific NLSE domain.
   
\par Our work is presented as follows: We begin by outlining the PINN method for solving nonlinear PDEs in Section 2.
In Section 3, we discuss the prediction of second-order positon solutions for a hierarchy of NLSEs using the PINN approach. In this section, we also conduct an error analysis between exact and predicted positon solutions for various systems, presenting the results graphically. Section 4 culminates with a concise summary of our key findings.
%
%
 \section{Methodology of Solving PDEs using PINN}
This section delves into the application of the PINN algorithm for tackling complex nonlinear PDEs. We begin by considering a general form of a complex PDE
 \begin{equation}
 	ir_t +  \mathcal{M}_r (r, r_x, r_{xx}, r_{xxx},....) = 0,~\label{general}
        x \in \Omega, t \in [t_0,T],
 \end{equation}
 where $r$ represents the complex wave function with space ($x$) and time ($t$) variables. $\mathcal{M}_r$ is the  general nonlinear function that consists of $r$ and its higher order derivatives with respect to $x$, $\Omega$ is a subset of ${\rm I\!R^D}$.  Complex differential equations cannot be solved by neural network directly.  Hence, we disintegrate the complex function $r(x,t)$ into real and imaginary parts as $r= u + i v$. Substituting this in Eq.~\eqref{general}, we can get the following two expressions, namely
 \begin{eqnarray}
 \begin{aligned}
 u_t + \mathcal{M}_u (u, u_x, u_{xx},..,v, v_x,....) & =&0,\\
 v_t + \mathcal{M}_v (v, v_x, v_{xx},..,u, u_x,....) & =&0.
 \label{split}
 \end{aligned}
 \end{eqnarray}
The residuals of PINN, $g_u (x,t)$ and $g_v(x,t)$, can be written as 
 \begin{eqnarray}
 \begin{aligned}
 	g_u &:=& u_t +  \mathcal{M}_u (u, u_x,,..,v, v_x,....) , \label{hu}\\
 	g_v &:=& v_t + \mathcal{M}_v (v, v_x,..,u, u_x,....) ,\label{hv}
  \end{aligned}
 \end{eqnarray}
where $\mathcal{M}_u$ and $\mathcal{M}_v$ are the nonlinear functional of $u$ and $v$ and their spatial derivatives. The functions $u(x,t)$ and $v(x,t)$ are to be determined by the deep neural network.  
\begin{figure*}[ht!]
\centering
\includegraphics[width=1\linewidth]{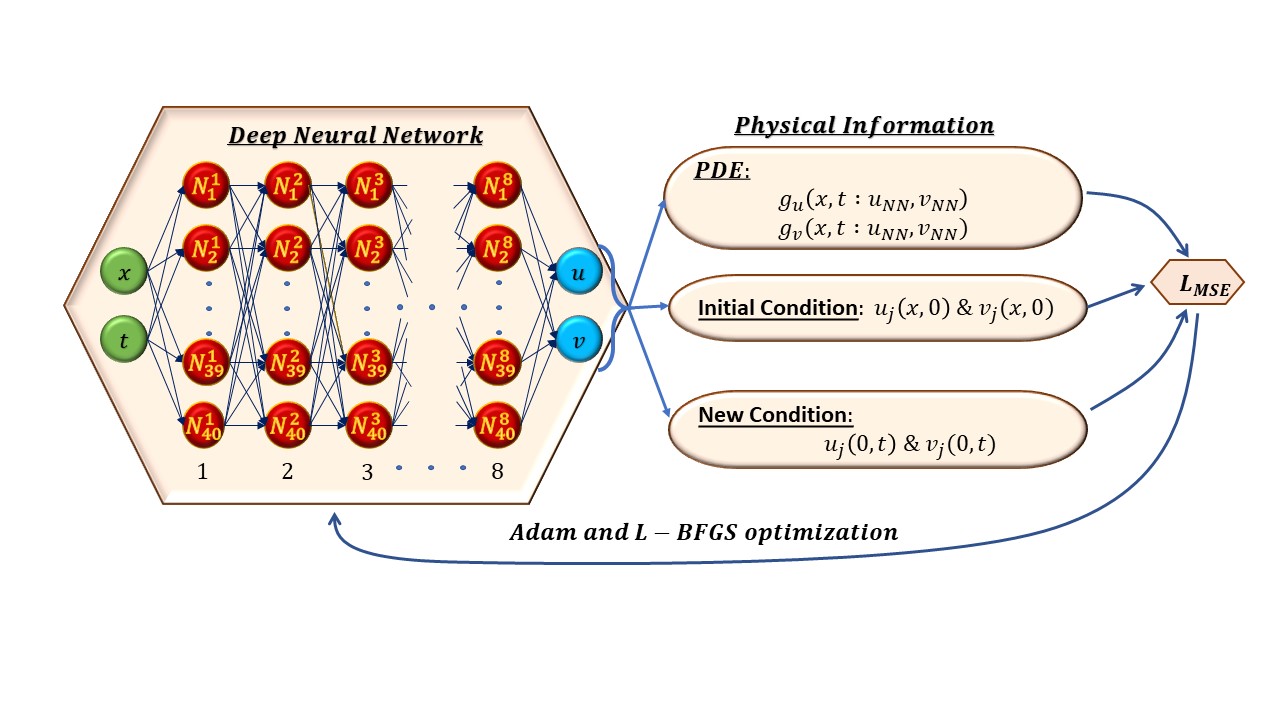}
	\caption{Graphic representation of PINN model.  The physical information inherent in the Physics-Informed Neural Network (PINN) is employed as the training loss function within the optimization problem.}
	\label{Network Model}
\end{figure*}
\par The PINN setup for solving the given PDE is given in Fig. {\ref{Network Model}}. $N_i^j$ in Fig. {\ref{Network Model}} represents the $i^{th}$ neuron of the $j^{th}$ layer. The output of every neuron takes the form
\begin{equation}
    \chi_l = \varsigma(w_l \chi_{l-1} + b_l). 
    \label{equ_nn}
\end{equation}
We denote the output of the $(l-1)$th layer as $\chi_{l-1}$. The activation function applied to this layer is $\varsigma$, and the weights and biases are represented by  $w_l$ and $b_l$, respectively. When $l = 0$, $\chi_0 = (x,t)$ represents the input layer of the network. Initially, weights and biases are randomly chosen using the Xavier initializer (or Glorot initializer) \cite{raissi-pinn}. Here, we use the hyperbolic tangent function, $tanh(.)$, as the activation function for every neuron.

The network is trained to find solutions, represented by $u(x,t)$ and $v(x,t)$, that minimize a loss function. This loss function combines the governing PDE along with the initial and boundary conditions. For training the model, we leverage data within the domain $x=0$. The network learns by minimizing the following mean squared error loss functions:
\begin{subequations}
\begin{equation}
MSE =\varrho_{IC} MSE_{IC} +\varrho_{NC} MSE_{NC} +\varrho_{E} MSE_E,\label{loss1}
\end{equation}
where \begin{eqnarray}
MSE_{IC} &=& \frac{1}{N_{IC}} \sum_{j=1}^{N_{IC}} |\hat{u}(x_0^j, 0) - u(x_0^j, 0)|^2  + |\hat{v}(x_0^j, 0)- v(x_0^j, 0)|^2,\\
MSE_{NC} &=& \frac{1}{N_{NC}} \sum_{j=1}^{N_{NC}} |\hat{u}(0,t_{NC}^j)- u(0,t_{NC}^j)|^2 + |\hat{v}(0,t_{NC}^j)- v(0,t_{NC}^j)|^2,\notag\\ \\
MSE_{E} &=& \frac{1}{N_E} \sum_{j=1}^{N_E}|g_u(x_E^j,t_E^j)|^2+|g_v(x_E^j,t_E^j)|^2.
\end{eqnarray}
\end{subequations}\label{mse1}
In the above, $\{x_0^j,u(x_0^j,t_0),v(x_0^j,t_0)\}_{j=1}^{N_{IC}}$ corresponds to the initial sampling points and $MSE_{IC}$ is the loss function for the initial data. In the proposed PINN algorithm, we have considered the initial condition as $t=0$, which typically characterize an initial value problem where the solution is determined by specifying the values at a single initial point. Boundary value problems typically involve specifying conditions at multiple points along the boundary of the domain. Unlike conventional PINN approaches that rely on Dirichlet boundary conditions, we exclusively utilize data points at a single reference point, $x=0$. This indicates that our model addresses an Initial Value Problem, where the solution is determined by providing initial conditions at a single point ($t=0$).  Typically, $x\in(-L,L)$ is used for boundary conditions. However, we deviate from the conventional approach by providing data at $x = 0$ instead of boundary conditions, which minimizes the loss function more rapidly for positon solutions. Thus, $\{t_{NC}^j,u(0,t_{NC}^j),v(0,t_{NC}^j)\}_{j=1}^{N_{NC}}$ correspond to the data points at $x = 0$, and $MSE_{NC}$ is the loss function arising from the condition $x = 0$. $\{x_E^j,t_E^j,g_u(x_E^j,t_E^j),g_v(x_E^j,t_E^j)\}_{j=1}^{N_{E}}$ are collocation points selected from the domain, and $MSE_{E}$ penalizes the given equations $g_u(x,t)$ and $g_v(x,t)$. We use the Latin Hypercube Sampling (LHS) strategy to obtain the collocation points for training the neural network \cite{raissi-pinn}. The spatial and temporal variables are discretized into $520$ and $400$ nodes, respectively. Therefore, we have a matrix size of $520 \times 400$ for the determined magnitude of density $|r(x,t)|^2$. $\varrho_{IC}$, $\varrho_{NC}$, and $\varrho_{E}$ are the weight correction coefficients for $MSE_{IC}$, $MSE_{NC}$, and $MSE_{E}$, respectively, which can be determined through: 
\begin{equation}
    \varrho_i = \frac{MSE_i}{\max({MSE_{IC}, MSE_{NC}, MSE_{E}})}; \quad i = IC,  NC, E.
\end{equation}
Initially, these coefficients are set to 1. It should be noted that these coefficients vary as the mean squared error (MSE) varies. We simulate the initial data using a specific number of iterations in Adam and L-BFGS optimization. Once training is completed, the PINN setup can predict the solution of the considered equation in the given space-time domain.
%
%
\section{Data driven positon solutions of a hierarchy of NLSE} 
In this section, we investigate second-order positon solutions of a family of nonlinear Schr\"{o}dinger equations (NLSEs) with higher-order nonlinearity using PINN approach. Specifically, we analyze the following nonlinear systems: (i) NLSE, (ii) third-order NLSE, (iii) fourth-order NLSE, (iv) fifth-order NLSE, (v) sixth-order NLSE, (vi) two coupled NLSE and (vii) two coupled Hirota equations. Since all these equations are complex, we split them as shown in (\ref{split}). For two-component systems, we split the functions as $r_1 = u_1 + iv_1$ and $r_2 = u_2 + iv_2$.

Our PINN architecture employs a single input layer, followed by eight hidden layers, and a single output layer. This configuration enables the network to predict the second-order positon solution of the governing equation.The input and output layers contain two neurons each, while each hidden layer consists of 40 neurons. We leverage the hyperbolic tangent function ($\tanh$) as the activation function within the network layers. The mean squared error loss function guiding the training process. We use the hyperbolic tangent function, as the activation function. The mean squared error loss function is formulated as presented in (\ref{loss1}).

In the following subsections, we describe how we utilize PINN to predict the positon solutions of the considered higher-order NLSEs and coupled equations.
\subsection{ Data driven second order positon solution of NLSE}
The NLSE plays a significant role in describing  various phenomena in plasma physics, biophysics and in nonlinear optics \cite{nls}. A widely recognized solution of the NLSE equation is solitons, which retain its shape after interaction with another soliton \cite{soliton}. Covering the wide capability of NLSE, abundant nonlinear wave solutions such as breathers, rogue waves, bright and dark solitons, to name a few, are constructed using various mathematical methods. As far as the utilization of PINN algorithm is concerned, in the literature, only soliton, breather and rogue wave solutions of the NLSE are simulated \cite{meiy-ml3,zhenya1}. While various solutions exist for the NLSE, this study concentrates on exploring the properties of positon solutions,
\begin{equation}
	 ir_t + r_{xx} + 2 |r|^2 r = 0, \label{nls}
\end{equation}
through PINN algorithm, where $r$ represents the complex wave envelope with $x$ and $t$ are taken as propagation and time variable, respectively. As the wave functions possess complex values, we decompose them into their real and imaginary components, $r(x,t) = u(x,t)+i v(x,t)$. Substituting this in \eqref{nls}, we get the real and imaginary parts of NLSE in the following form  
\begin{eqnarray}
	\begin{aligned}
	g_u(x,t) &=& 2 u^3 + 2 u  v^2 - v_{t} + u_{xx},  \\
	g_v(x,t) &=& 2 v^3 + 2 u^2 v +  u_{t} + v_{xx}.
		\end{aligned}
\end{eqnarray}
\par The exact positon solutions of the NLSE can be found in reference \cite{w1}, where they were determined using the generalized DT method. We present the explicit form of the positon solution of \eqref{nls} in the Appendix, see Eq.\eqref{nls1}. This solution is utilized to obtain initial and boundary data for training the PINN. Our simulations are conducted within a spatial and temporal domain of $[-10,10]$ for both position ($x$) and time ($t$).
We initialize the data $r(x,0)$, by choosing $N_{IC}=100$ random samples. Additionally, $N_{NC}=200$ sample points are selected specifically at $x = 0$ and $N_E = 10000$ random points are drawn within the region $-10 \leq x \leq 10$ and $-10 \leq t \leq 10$. To minimize the loss function, we perform 80000 iterations in Adam optimization. 

\begin{figure}[ht!]
\includegraphics[width=1\linewidth]{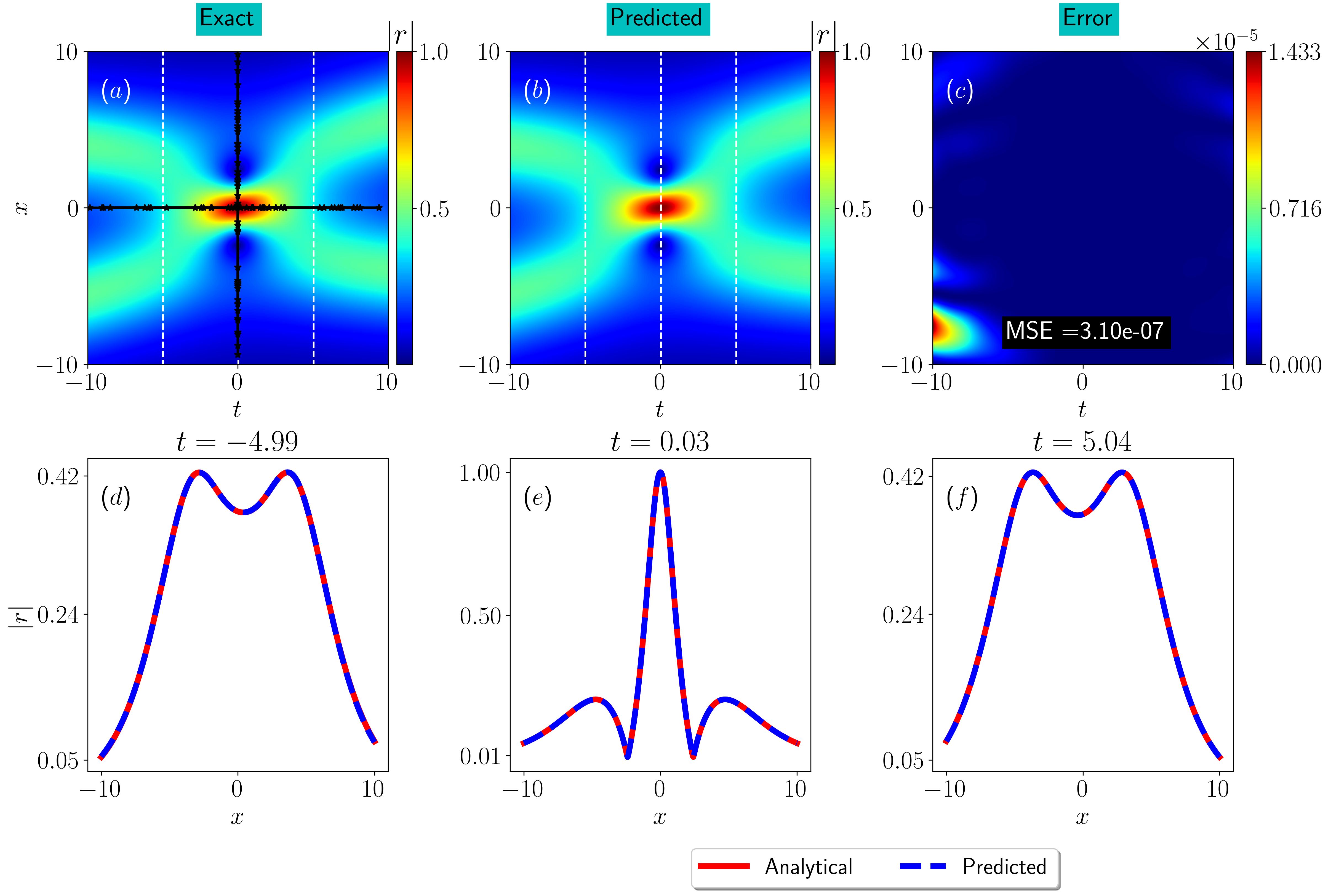}\caption{Predicted and exact solution of the second order positon of NLSE using PINN approach. (a) Contour profile of the analytical second order positon solution with parameter value $\lambda = 0.01+ 0.25 i$, (b) The predicted second order positon solution by PINN. (c) Mean squared error plot between  the analytical and predicted solution. (d)-(f) are the various time snapshots of predicted and exact solution respectively. The black line in (a) represents the data points taken at initial and new conditions, respectively.\label{2}}
\end{figure}

Figure \ref{2}(a) visualizes the second-order positon solution obtained using the generalized DT method, depicted as a contour plot. The predicted solution of the model accurately matches the second-order positon solution of Eq.\eqref{nls}, as shown in Fig. \ref{2}(b), and the mean squared error calculated for the model is in the order of $10^{-7}$, as seen in Fig. \ref{2}(c). Furthermore, snapshots at various time regimes confirm that the positon solution predicted through PINN is highly reliable.

%
%
\subsection{Data-driven positon solution of third order NLSE}
When exploring the generation of ultrashort pulses in the sub-picosecond or femtosecond regime, a crucial equation comes into play: the Hirota equation, also known as the third-order NLSE. This equation effectively describes the propagation dynamics of such pulses by incorporating higher-order effects beyond the standard NLSE. The Hirota equation can be represented as \cite{hirota}:
\begin{equation}
	ir_t + r_{xx} + 2 |r|^2 r - i \alpha \left( r_{xxx} + 6 r_x|r|^2 \right) =0,\label{hirota}
\end{equation}
 where $r$ represents the complex wave function and $\alpha$ is the nonlinear and dispersion parameter. In the context of ultra-short laser pulse propagation, it is crucial to consider higher-order dispersive and nonlinear effects, including phenomena like self-frequency shift, third-order dispersion, and self-steepening effects. Researchers have employed a variety of techniques to discover localized solutions of the Hirota equation. These methods include the inverse scattering transform, Hirota bilinear method, and Darboux transformation. The smooth positon solutions of the third-order NLSE can be found in references \cite{w1,w2}. Soliton, rogue waves, and breathers of the third-order NLSE/Hirota equation have also been investigated using deep learning methods \cite{tonls}. For our investigations, we set $\alpha=1$, $r = u(x,t) + iv(x,t)$, and split the Hirota equation into two real equations
\begin{eqnarray}
	\begin{aligned}
g_u (x,t) :=&  2 u^3 + 2 u v^2 - v_{t} + 6 u^2  v_{x} + 6 v^2  v_{x} + u_{xx} +  v_{xxx}, \\
g_v (x,t) :=&  2  u^2 v + 2  v^3 + v_{t} + v_{xx} - 6 u^2 v_{x}- 6 v^2 u_{x} - u_{xxx}.
	\end{aligned}
\end{eqnarray}
\par We provide the actual solution of the second-order positon solution of the Hirota equation with appropriate parameter values in the Appendix, see Eq.\eqref{hirota1}. This explicit solution is used to generate initial and boundary data for the PINN to train the solution. The network setup is similar to the NLSE case, with the mean squared error function (\ref{loss1}) calculated accordingly. The range for the space and time variables is taken as $x,t\in[-10,10]$ and sample points are chosen similar to the NLSE case.

\begin{figure}[ht!]
\includegraphics[width=1\linewidth]{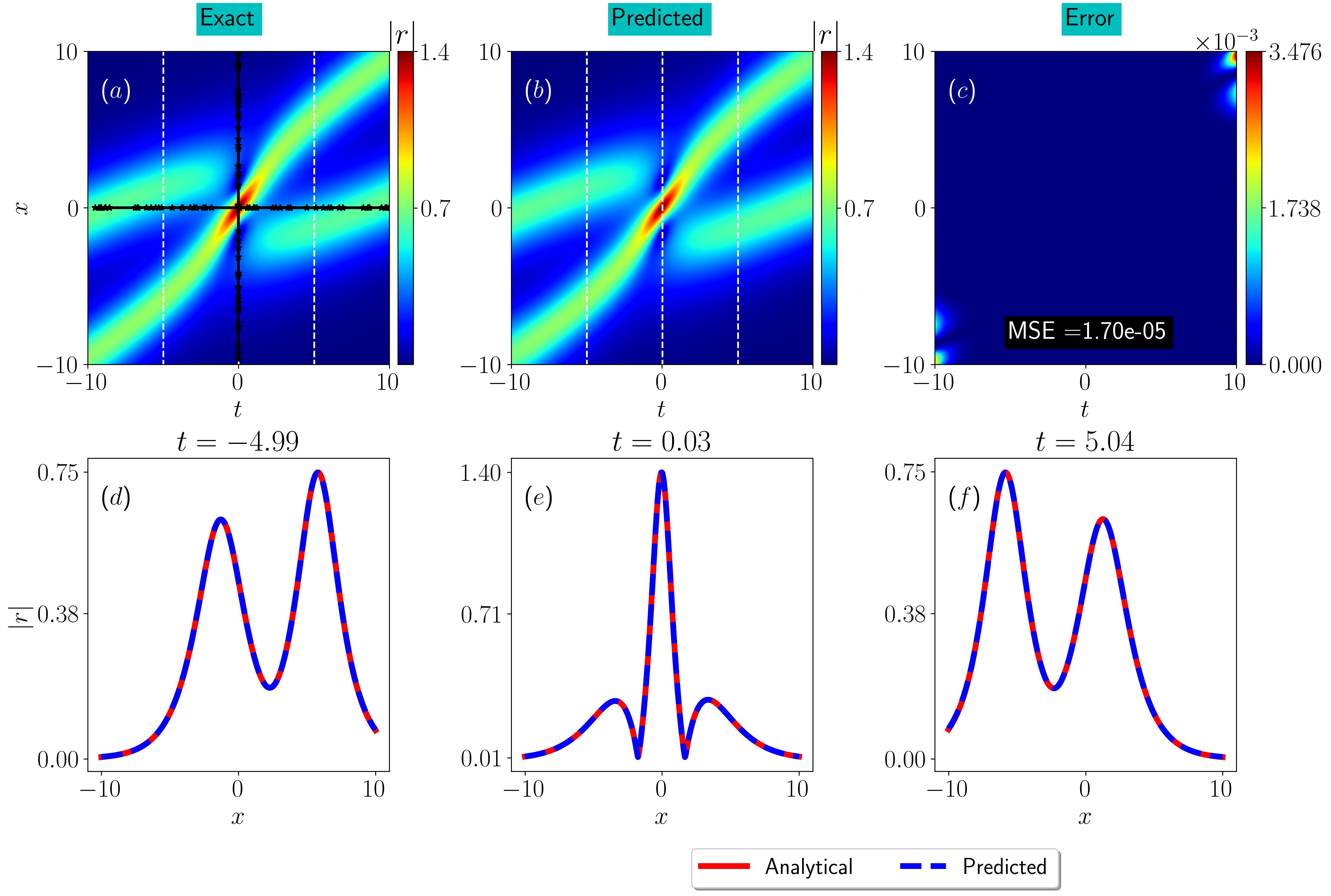}
	\caption{Predicted and exact solution of the second order positon of Hirota equation using PINN approach. (a) Contour profile of the analytical second order positon solution with parameter value $\lambda = 0.3+ 0.5 i$. (b) The predicted second order positon solution by PINN. (c) Mean squared error plot between the analytical and predicted solution. (d) and (f) are the various time snapshots of predicted and exact solution respectively. The black line in (a) represents the data points taken at initial and new conditions, respectively.\label{3}}
\end{figure}

After completing $70000$ iterations in Adam optimization, the PINN is able to predict the second-order positon solution of the Hirota equation with high accuracy. Figure~\ref{3}(a) depicts the exact second-order positon solution, while Fig. \ref{3}(b) shows the second-order positon solution predicted through the PINN algorithm. Figure \ref{3}(c) depicts the squared error between the predicted and exact outcomes of the second-order position. A comparison between the predicted and exact positon solutions is computed and plotted in Figs. \ref{3}(d)-(f) for different time regimes, validating the superiority of PINN.      
%
%
\subsection{Data driven positon solution of fourth order NLSE }
A higher order NLSE with fourth-order nonlinearity is proposed in \cite{lpd} in the context of  Heisenberg ferromagnetic spin chain.  This equation has also been used to study the propagation of periodic ultrashort pulses in optical fibers. Rogue waves, breathers, positons, bright, dark, kink and optical solitons are constructed for this fourth order NLSE in \cite{lpd-sol}. In \cite{tonls}, it has been stated that the PINN algorithm fails to predict solutions when the order of the differential equation is higher. In this paper, we  successfully predict and analyze the positon solution of the following fourth order NLSE,  
\begin{eqnarray}
 ir_t + r_{xx} + 2 |r|^2 r + \gamma ( r_{xxxx} + 8|r|^2r_{xx}+ 2r^2 r_{xx}\notag \\  + 4 r |r_x|^2 + 6 r_x^2 r + 6|r|^4 r )  = 0,\label{lpd}
\end{eqnarray}
using PINN method. Here, $r$ is the complex wave function and $\gamma$ is the coefficient of the fourth order nonlinear parameter.  To determine the solution of fourth order NLSE (\ref{lpd}) using PINN method, we first split the wave envelope into real and imaginary parts by substituting $r= u + i v$ and we choose $\gamma=1$ to illustrate the results in a simple manner.  Hence Eq. (\ref{lpd}) becomes
\begin{eqnarray}
	\begin{aligned}
	g_u (x,t) &= 2 u^3 + 6 u^5 + 2 u v^2 + 12  u^3  v^2 + 6  u  v^4 - v_{t }+ 10 u v_{x}^2 
	+ 12 v u_{x}  v_{x} \\&- 2 u  v_{x}^2 + u_{xx}  + 10  u^2  u_{xx} +  6  v^2  u_{xx} + 4  u v v_{xx }+ u_{xxxx}, \\
	g_v (x,t) &= 2 u^2 v + 6  u^4 v + 2 v^3 + 12 u^2  v^3 + 6 v^5 + u_t- 2 v u_x^2 + 12 u u_x  v_x \\&  + 10  v  v_x^2 + 4 u v  u_{xx}  + v_{xx}+ 6  u^2 v_{xx} + 10 v^2 v_{xx} + v_{xxxx}.
\end{aligned}
\end{eqnarray} \label{lpd_uv}
\begin{figure}[ht!]
\centering
\includegraphics[width=1\linewidth]{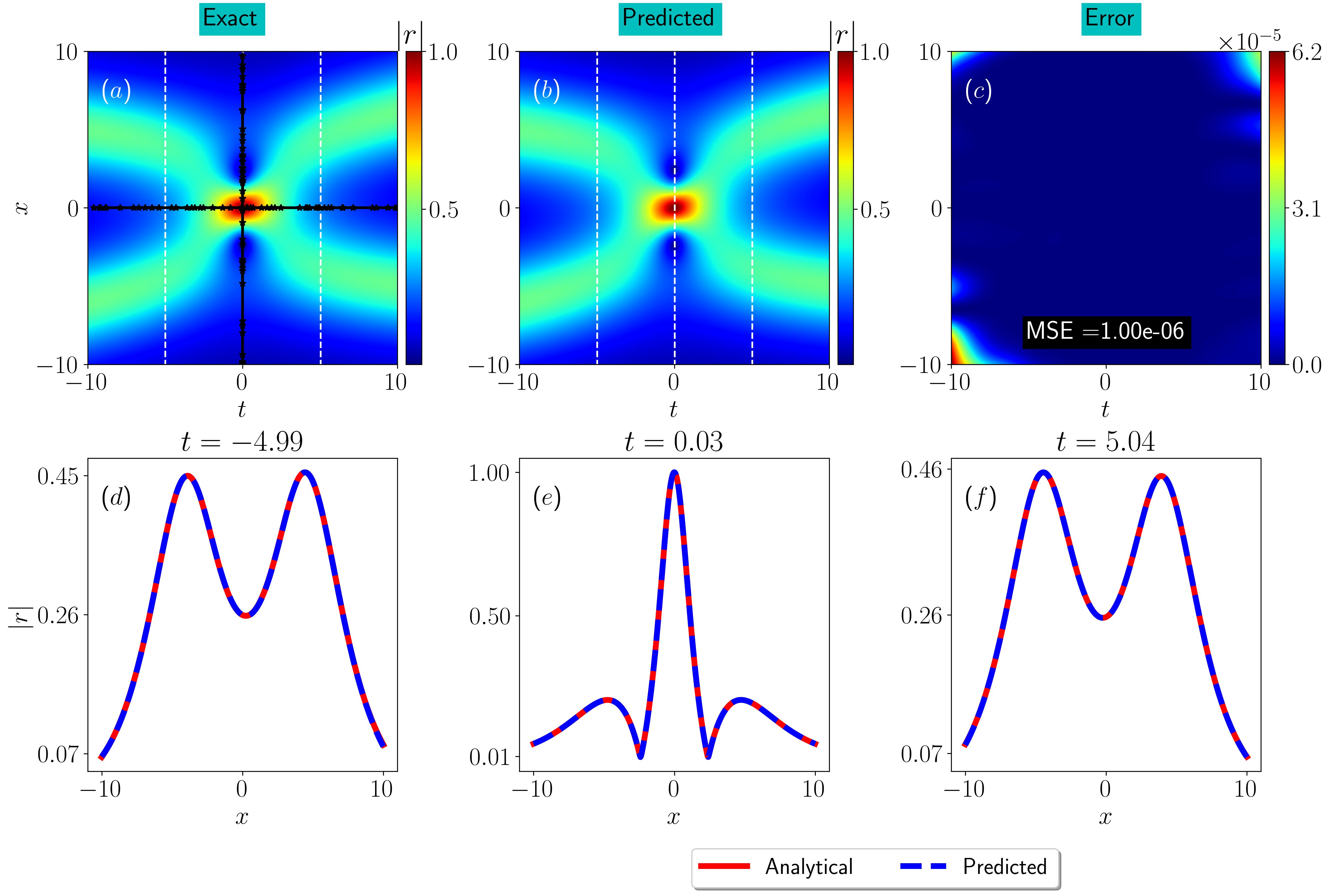}
\caption{Predicted and exact solution of the second order positon of fourth order NLSE using PINN approach. (a) Contour profile of  the analytical second order positon solution with parameter value $\lambda = 0.01+ 0.25 i$. (b) The predicted second order positon solution by PINN. (c) Mean squared error plot between the analytical and predicted solution. (d) and (f) are the various time snapshots of predicted and exact solution respectively. The black line in (a) represents the data points taken at initial and new conditions, respectively.}
\label{4}
\end{figure}
The exact second-order positon solution of the fourth-order NLSE \eqref{lpd} is presented in the Appendix (see Eq.\eqref{lpd1}). As previously explained, initial and boundary data are generated from the explicit solution. The PINN setup is similar to the previous two cases. Here, we select the spatial and time variables in the range $x\in[-10,10]$, $t\in[-10,10]$, and sample points are chosen similarly to the previous two cases.

After 50000 iterations in Adam optimization, the PINN is able to produce the positon solution of Eq.\eqref{lpd}. The output of the model is illustrated in Fig. \ref{4}. The exact positon solution diagram of \eqref{lpd} is given in Fig. \ref{4}(a), while Fig. \ref{4}(b) represents the corresponding contour plot of the positon solution. It is evident that the model is able to accurately predict the positon solution of Eq.\eqref{lpd}. The mean squared error is on the order of $10^{-6}$, and the error plot is shown in Fig. \ref{4}(c).

%
\par We also consider another form of NLSE in our study, which has both cubic and quartic nonlinearities, that is 
\begin{eqnarray}
\begin{aligned}
ir_t + r_{xx} + 2 |r|^2 r- i \alpha \left( r_{xxx} + 6 r_x|r|^2 \right)+ \gamma \big( r_{xxxx}+ 8|r|^2r_{xx} \\  + 2r^2 r_{xx} + 4 r |r_x|^2 + 6 r_x^2 r + 6|r|^4 r \big)  = 0,\label{enls}
\end{aligned}
\end{eqnarray}
where the parameters $\alpha$ and $\gamma$ are associated with third and fourth-order dispersion and the nonlinear term, respectively. The function $r(x,t)$ represents the wave envelope of the system. This equation is a generalized version of the higher-order NLSE, where when the nonlinear parameters $\alpha,\gamma$ are zero, the equation reduces to the conventional NLSE. Similarly, when $\alpha\neq0,\gamma=0$ or $\alpha=0,~\gamma\neq0$, the equation takes the form of the Hirota equation or the fourth-order NLSE, respectively. In earlier work \cite{w2}, higher-order smooth positon and breather positon solutions for Eq.\eqref{enls} were constructed.

\begin{figure}[ht!]
\centering
\includegraphics[width=1\linewidth]{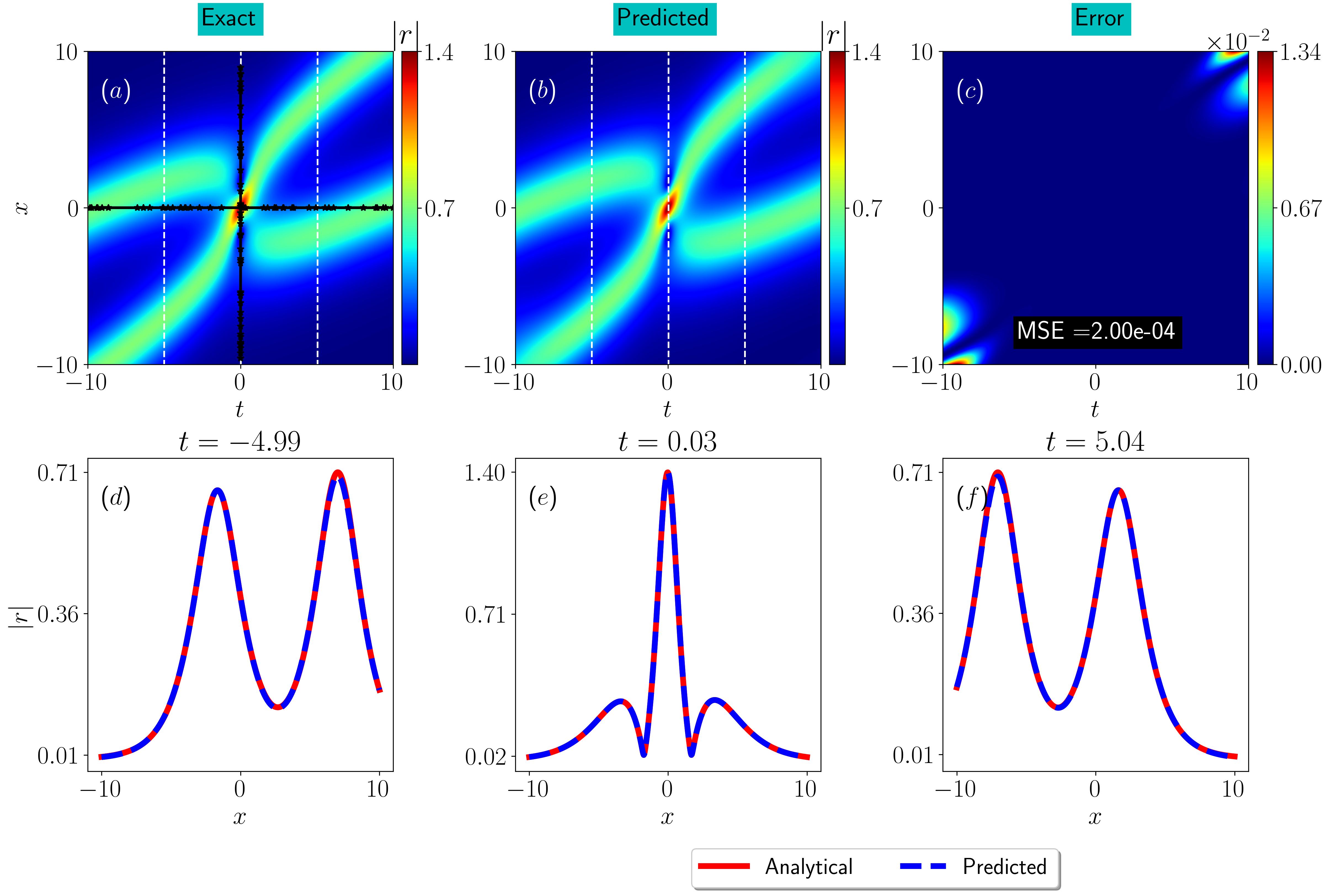}
\caption{Predicted and exact solution of the second order positon of extended NLSE using PINN approach. (a) Contour profile of the analytical second order positon solution with parameter value $\lambda = 0.01+0.35 i$. (b) The predicted second order positon solution by PINN. (c) Mean squared error plot between the analytical and predicted solution. (d) and (f) are the various time snapshots of predicted and exact solution respectively. The black line in (a) represents the data points taken at initial and new conditions, respectively.}
\label{5}
\end{figure}

To predict these smooth positon solutions of \eqref{enls}, we train the PINN model with the same setup used for the fourth-order NLSE \eqref{lpd}. In this case, we set the parameter values as $\alpha,~\gamma=1$ with the spatial and time range as $x, t\in[-10,10]$. The predicted results are shown in Fig. \ref{5}. The PINN model (Fig. \ref{5}(b)) has provided more accurate results with a low mean squared error value of the order of $10^{-4}$, as confirmed by Figs. \ref{5} (c)-(f). The PINN method has effectively derived the positon solution for Eq. \eqref{enls}, yielding a predicted solution that closely aligns with the exact solution.

%
%
\subsection{Data driven positon solution of a fifth order NLSE}
Next, we investigate a fifth order NLSE which narrates the one-dimensional anisotropic Heisenberg ferromagnetic spin chain \cite{fnls}. In the literature, solitons, Akhmediev breathers, Kuznetsov–Ma solitons and rogue waves for this fifth order NLSE are studied in \cite{fnls1,5nls-rw,5nls-sol}. The positon solutions are also constructed in \cite{fnls} for the fifth order NLSE. To study these positon solutions via PINN approach, we pick up the following fifth order NLSE   
\begin{eqnarray}
 ir_t + r_{xx} + 2 |r|^2 r- i \delta( r_{xxxxx}+ 10|r|^2
r_{xxx} + 30|r|^4 r_x +10 r r^{*}_{x} r_{xx}\notag\\+ 10 r r_x r^*_{xx}
 + 20 r^{*} r_x r_{xx} + 10 r^2_{x} r^{*}_{x}) = 0, \label{5nls}
\end{eqnarray}

where $*$ denotes complex conjugate and $\delta$ is the arbitrary real parameter. In this section, we investigate the positon solution of the fifth order NLSE through PINN algorithm.  Since $r$ is the complex valued function, we segregate the equation into real and imaginary parts by choosing the function $r=u+iv$ and fix the parameter $\delta = 0.15$. The resultant action yields the following two expressions, that is 
\begin{subequations}
\begin{eqnarray}
	g_u (x,t) &=& 2 u^3 + 2 u  v^2 + u_{xx} + 30\delta u^4v_x + 30\delta  v^4v_x  + 60\delta u^2v^2v_x
          + 10\delta u_x^2v_x \notag \\&&+ 10\delta v_x^3 - v_t  + 20\delta uv_x u_{xx} + 20\delta uu_x v_{xx}
          + 40\delta vv_xv_{xx} \notag \\&& + 10\delta u^2v_{xxx} + 10\delta v^2v_{xxx} + \delta v_{xxxxx}, \\
	g_v (x,t) &=& 2  u^2  v + 2  v^3  + v_{xx} - 30\delta u^4u_x - 30\delta v^4u_x- 60\delta u^2v^2u_x 
          - 10\delta u_x^3 \notag \\&&- 10\delta u_x v_x^2 + u_t  - 40\delta u u_x u_{xx} - 20\delta v v_x u_{xx}
          - 20\delta v u_x v_{xx} \notag \\&&- 10\delta u^2u_{xxx} - 10\delta v^2u_{xxx} - \delta u_{xxxxx}.
\end{eqnarray} \label{fifth_uv}
\end{subequations}
Utilizing the gDT method, we derived the second-order positon solution for Eq.\eqref{5nls}, and the exact expression of the constructed solution is provided in the Appendix, see Eq.\eqref{fnls1}. From this solution, we extracted the initial and boundary data points. The PINN setup was configured as in the previous cases, with the spatial and time range set to $x, t \in[-10,10]$. After implementing 30000 iterations in Adam optimization, the PINN network predicted the second-order positon solution of the fifth-order NLSE \eqref{5nls}.

\begin{figure}[ht!]\centering\includegraphics[width=1\linewidth]{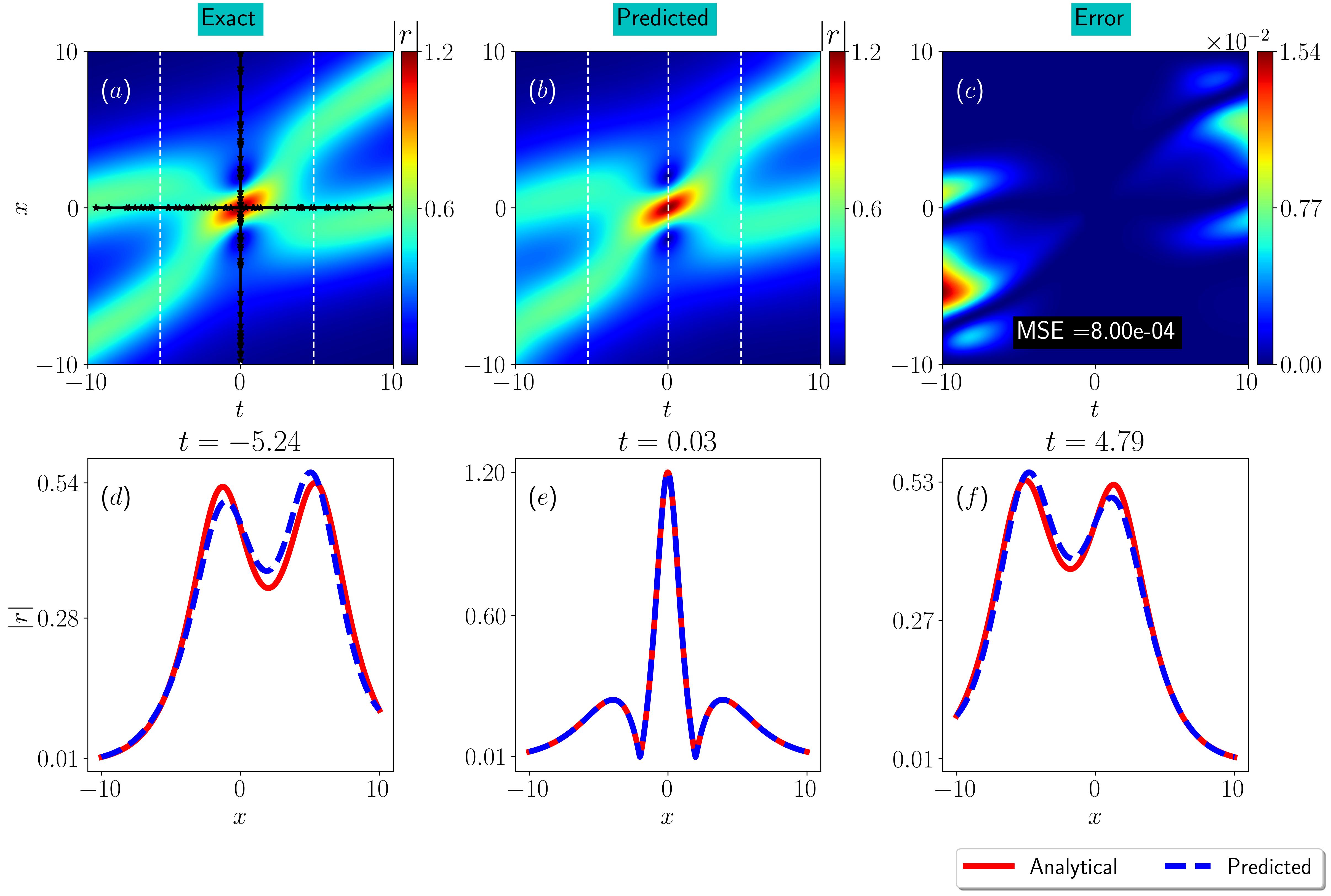}\caption{Predicted and exact solution of the second order positon of fifth order NLSE using PINN approach. (a) Contour profile of the analytical second order positon solution of two components $r_1$ and $r_2$ with parameter value $\lambda = 0.1+ 0.3 i$ and $\delta  = 0.15$. (b) The predicted second order positon solution by PINN. (c) Mean squared error plot between the analytical and predicted solution. (d) and (f) are the various time snapshots of predicted and exact solution respectively. The black line in (a) represents the data points taken at initial and new conditions, respectively.}\label{7}\end{figure}

Figures \ref{7} (a) and \ref{7} (b) depict the exact and predicted solutions of the fifth-order NLSE, respectively. The mean squared error value was significantly reduced to the order of $10^{-4}$, as demonstrated in the error plot, see Fig. \ref{7} (c). Additionally, a comparison of time snapshots at various regimes between the exact and predicted results is shown and plotted in Figs. \ref{7} (d)-(f), further confirming the high precision of PINN.

%
%
\subsection{Data driven positon solution of a sixth order NLSE}
\begin{figure}[ht!]
\centering\includegraphics[width=1\linewidth]{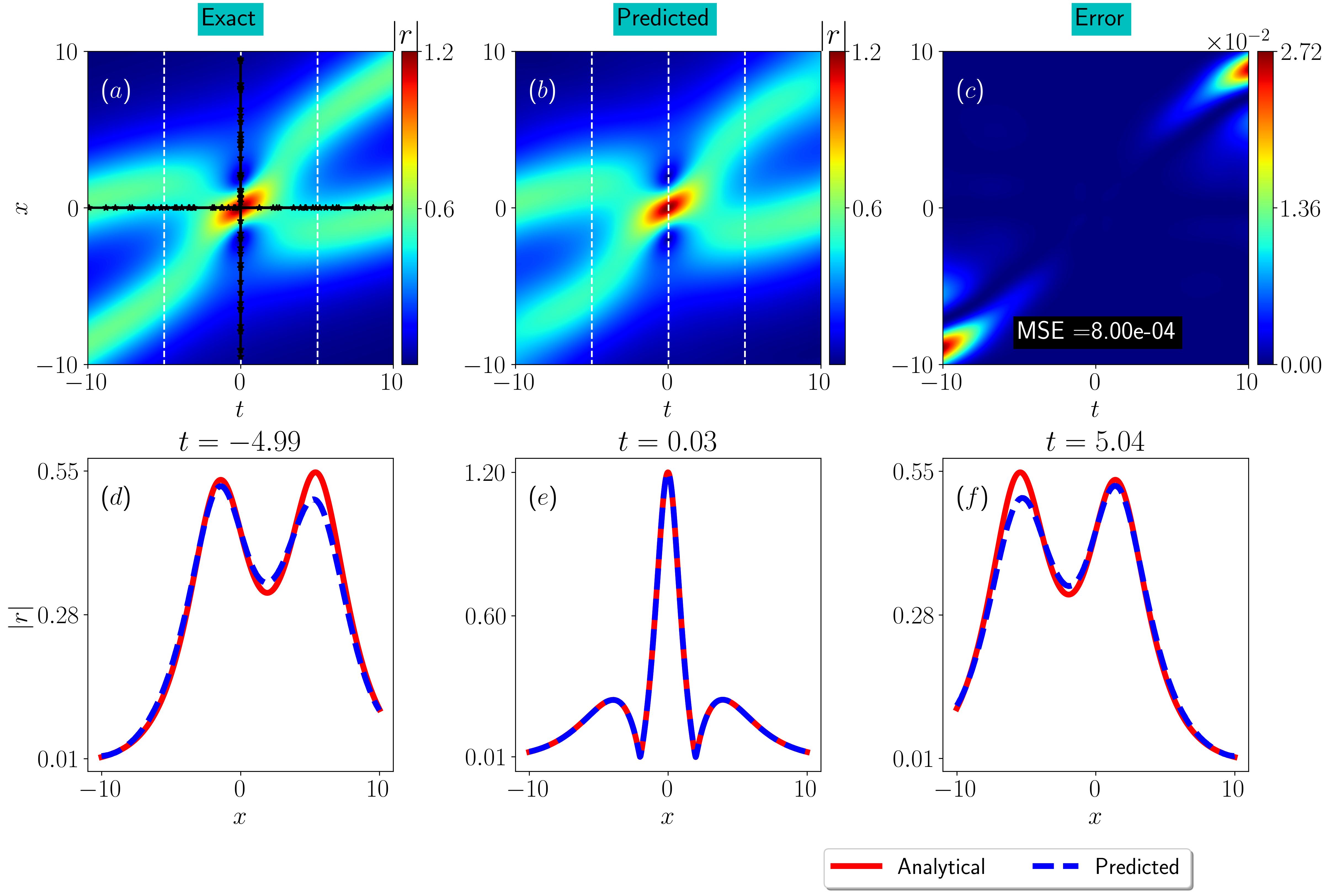}\caption{Predicted and exact solution of the second order positon of sixth-order NLSE using PINN approach. (a) Contour profile of the analytical second order positon solution of two components $r_1$ and $r_2$ with parameter value $\lambda = 0.1+ 0.3 i$ and $\xi  = 0.15$. (b) The predicted second order positon solution by PINN. (c) Mean squared error plot between the analytical and predicted solution. (d)-(f) are the various time snapshots of predicted and exact solutions. The black line in (a) represents the data points taken at initial and new conditions, respectively.}\label{7a}\end{figure}

The sixth order NLSE is one of the extended version of the basic NLSE to incorporate the effects of dispersion at higher orders. This becomes crucial when considering shorter-duration pulses propagating in fibers or other media. The sextic NLSE, shown below, explicitly accounts for these higher-order effects, particularly the sextic-order dispersion term $\xi r_{xxxxxx}$, where $\xi$ is a real parameter
\begin{eqnarray}
  &i r_{t} + r_{xx} + 2|r|^{2} r + \xi \big(r_{xxxxxx} + [60 r^{*} |r_x|^{2} + 50(r^{*})^{2} r_{xx}+ 2 r^{*}_{xxxx} ]r^{2} \notag \\& + r [12 r^{*} r_{xxxx}+ 8 r_x r^{*}_{xxx} + 22 |r_{xx}|^{2} ]+
  r [18 r_{xxx} r^{*}_x + 70 (r^{*})^{2} r^2_x ] + 20 (r_x )^{2} r^{*}_{xx} \notag \\&+ 10 r_x [5 r_{xx} r^{*}_x + 3 r^{*} r_{xxx}] + 20 r^{*} r^{2}_{xx}+ 10 r^{3} [(r^{*}_x )^{2} + 2 r^{*} r_{xx}^{*} ]+ 20 r|r|^{6}\big) = 0. \label{6nls}\notag \\
\end{eqnarray}
The sextic NLS equation is known to admit various solutions, including Lax pair, solitons, breathers, periodic solutions, and rogue waves. \cite{6nls1}. Breather-to-soliton transitions and the interactions between different types of nonlinear waves for Eq.~\eqref{6nls} are explored in \cite{6nls2}. Equation \eqref{6nls} provides a more comprehensive framework for modeling and analyzing the behaviour of nonlinear waves in various physical systems, offering insights into complex wave dynamics. In this work, we have derived the second order positon solution for the sextic NLSE \eqref{6nls} using the GDT method (detailed calculation of GDT will be published elsewhere). This study investigates the application of PINNs to predict second-order positon solutions within this framework, hence we present only the explicit solution in the Appendix \eqref{6nls1} for brevity. For that, We extract the real and imaginary parts from the complex eigenfunction $r=u + i v$, yielding:
\begin{eqnarray}
    g_u(x,t) && =  2u^3 + 20\xi u^{7} + 2u v^{2} + 60 \xi u^{5} v^2 + 60\xi u^3 v^4 + 20\xi  u v^6 - v_t + u_{xx}\notag \\&&
          + 140\xi u^3 u_{x}^2
          + 100\xi u v^2 u_{x}^2 + 200 \xi u^2 v u_x v_x + 120\xi v^3 u_x v_x - 20\xi u^3 v_{x}^2 \notag \\&&
          + 20\xi u v^2 v_x^2  + 70 \xi u^4 u_{xx} +100\xi u^2 v^2 u_{xx} + 30\xi v^4 u_{xx} 
          + 70\xi u_x^2 u_{xx}\notag \\&& + 30\xi v_x^2 u_{xx} + 42\xi u u_{xx}^2 + 40\xi u^3 v v_{xx} 
          + 40\xi u v^3 v_{xx} + 40\xi u_x v_x v_{xx} \notag \\&&+ 40\xi v u_{xx} v_{xx} + 2\xi u v_{xx}^2
          + 56\xi u u_x u_{xxx} + 40\xi v v_x u_{xxx} + 20\xi v u_x v_{xxx}\notag \\&& - 4\xi u v_x v_{xxx}
          + 14\xi u^2 u_{xxxx} + 10\xi v^2 u_{xxxx} + 4\xi u v v_{xxxx} + \xi u_{xxxxxx}\notag \\
    g_v(x,t) & &=  2 u^2 v + 20\xi u^6 v + 2 v^3 + 60\xi u^4 v^3 + 60 \xi u^2 v^5 + 20 \xi v^7 + u_t+ v_{xx}\notag \\&&
          + 20\xi u^2 v u_x^{2} - 20\xi v^3 u_x^2 + 120 \xi u^3 u_x v_x + 200\xi u v^2 u_x v_x 
          + 100\xi u^2 v v_x^2 \notag \\&&+ 140\xi v^3 v_x^2 + 40\xi u^3 v u_{xx} + 40\xi u v^3 u_{xx} 
          + 40 \xi u_x v_x u_{xx} + 2\xi v u_{xx}^2  \notag \\&&+30 \xi u^4 v_{xx} + 100 \xi u^2 v^2 v_{xx}
          + 70 \xi v^4 v_{xx} + 30\xi u_x^2 v_{xx} + 70 \xi v_x^2 v_{xx} \notag \\&&+ 40\xi u u_{xx} v_{xx}
          + 42\xi v v_{xx}^2 - 4\xi v u_x u_{xxx} + 20\xi u v_x u_{xxx} + 40 \xi u u_x v_{xxx}\notag \\&& 
          + 56\xi v v_x v_{xxx} + 4\xi u v u_{xxxx} + 10\xi u^2 v_{xxxx} + 14\xi v^2 v_{xxxx}
          + \xi v_{xxxxxx}\notag \\
\end{eqnarray}
The initial and boundary data points were extracted, and the PINN setup was constructed following the procedures used in prior cases, with the spatial and time range defined as $x, t \in[-10,10]$. After executing $50,000$ iterations in Adam optimization, the PINN successfully predicted the second-order positon solution of the sixth-order NLSE \eqref{6nls}. Figures \ref{7a} (a) and \ref{7a} (b) depict the exact and predicted solutions of the sixth-order NLSE. The mean squared error value significantly decreased to the order of $10^{-4}$, as demonstrated in the error plot shown in Fig. \ref{7a} (c). Additionally, the time snapshots at various regimes between the exact and predicted results were compared and plotted in Figs. \ref{7a} (d)-(f), confirming the high precision of the PINN.
%
%
\begin{figure}[ht!]
\centering
\includegraphics[height=10cm,width=1\linewidth]{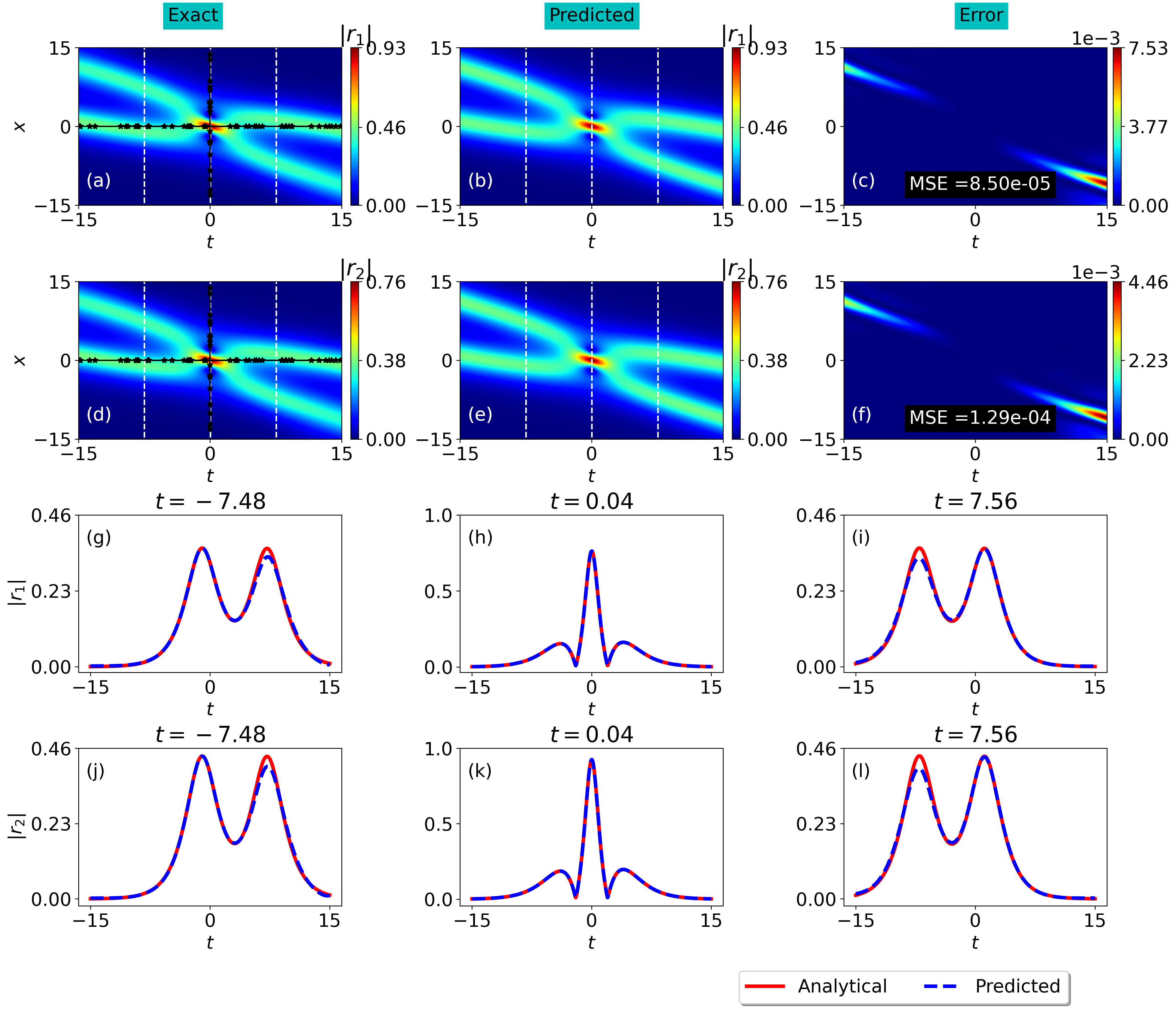}
\caption{Predicted and exact solution of second order positon of coupled NLSE using PINN approach. (a),(d) Contour profile of analytical second order positon solution of two components $r_1$ and $r_2$ with parameter values $\lambda = 0.1+ 0.3 i$, $s_1 =1.05 ,s_2 = 0.7, s_3 = 0.85 $. (b) and (e) are the predicted second order positon solution by PINN. (c),(f) Mean squared error plot between analytical and predicted solution. (g)-(l) are the various time snapshots of predicted and exact solution respectively. The exact region of snapshots is denoted as white dotted lines in (a), (b), (d) and (e). The black line in (a) and (d) represents the data points taken at initial and new conditions, respectively.}
\label{8}
\end{figure}

\subsection{Data driven positon solution of coupled NLSE}
In the previous sub-secs., 3.1-3.4, we have studied the positon solution for the family of scalar NLSEs through PINN approach. In this section, we consider a  two component NLSE which is essential for studying optical pulse propagation in birefringent optical fibers \cite{manakov,bec1}. Positon solution of Eq.\eqref{cnls} has been discussed in \cite{cnls} and the authors have also studied the interaction between soliton and positon solutions for the following coupled NLSEs: 
\begin{eqnarray}
\begin{aligned}
i r_{1t} + r_{1xx} +  2 ( |r_1|^2 + |r_2|^2)r_1 = 0,\\
i r_{2t} + r_{2xx} +  2 ( |r_1|^2 + |r_2|^2)r_2 = 0.\label{cnls}
\end{aligned}
\end{eqnarray}
Here $r_1$ and $r_2$ are the wave envelope functions. For the PINN approach, we split these complex wave functions into real and imaginary parts by choosing $r_{1,2} = (u_{1,2} + i v_{1,2})$ in Eq.\eqref{cnls}. The resultant expressions read
\begin{eqnarray}
g_{u_{1}}(x,t) &=& 2 u_1^3 + 2 u_1 u_2^2 + 2  u_1  v_1^2 + 2 u_1 v_2^2 - v_{1t} + u_{1xx},\notag\\
g_{v_{1}}(x,t) &=& 2  u_1^2  v_1 + 2  u_1^2  v_1 + 2 v_1^3 + 2 v_1 v_2^2 + u_{1t} + v_{1xx},\notag\\
g_{u_{2}}(x,t) &=&  2 u_2^3 + 2 u_2 u_1^2 + 2 u_2 v_2^2 + 2 u_2 v_1^2 - v_{2t} + u_{2xx},\notag\\
g_{v_{2}}(x,t) &=& 2 u_2^2 v_2 + 2 u_1^2 v_2 + 2 v_2^3 + 2 v_2 v_1^2 + u_{2t} + v_{2xx}.\notag\\
\end{eqnarray}
\par For the two component system, the mean squared error loss functions are calculated through the modified expressions, 
\begin{subequations}
\begin{eqnarray}
MSE_{IC} &=&\frac{1}{N_{int}} \sum_{j=1}^{N_{int}} \bigg( |\hat{u_1}(x_0^j, 0) - u_1(x_0^j, 0)|^2    + |\hat{v_1}(x_0^j, 0)- v_1(x_0^j, 0)|^2 \notag \\&&+ |\hat{u_2}(x_0^j, 0)  - u_2(x_0^j, 0)|^2+ |\hat{v_2}(x_0^j, 0) - v_2(x_0^j, 0)|^2\bigg), \\
	MSE_{NC} &=& \frac{1}{N_{NC}} \sum_{j=1}^{N_{NC}} \bigg(|\hat{u_1}(0,t_{NC}^j) - u_1(0,t_{NC}^j)|^2 + |\hat{v_1}(0,t_{NC}^j) - v_1(0,t_{NC}^j)|^2 \notag \\&& +  |\hat{u_2}(0,t_{NC}^j)  - u_2(0,t_{NC}^j)|^2+ |\hat{v_2}(0,t_{NC}^j) - v_2(0,t_{NC}^j)|^2\bigg),\\
	MSE_{E} &=& \frac{1}{N_E} \sum_{j=1}^{N_E}\bigg(|g_{u_1}(x_E^j,t_E^j)|^2+|g_{v_1}(x_E^j,t_E^j)|^2 + |g_{u_2}(x_E^j,t_E^j)|^2 \notag \\&& +|g_{v_2}(x_E^j,t_E^j)|^2 \bigg).
\end{eqnarray}\label{mse2}
\end{subequations}
\par The second order positon solution for the coupled NLSE was constructed in \cite{cnls} using generalized DT method. The explicit expressions of both the components $r_1$ and $r_2$ are given in Appendix, see Eq.\eqref{cnls1}. The same setup of PINN was implemented with $4$ neurons in the output layer for the two component system. The spatial and time variable are taken as $x, t\in[-15,15]$ and we take $30000$ iterations in Adam optimization.  Using this PINN algorithm, we predict the second order positon solution of the coupled NLSE \eqref{cnls}. The predicted outcomes are given in Fig. \ref{8}. The exact and predicted results of the components $r_1$ and $r_2$ are shown in Figs. \ref{8}(a), (b), (d) and (e), respectively. The error plot between exact and predicted solutions is illustrated in Fig. \ref{8} (c) and (f) with mean squared error in the order of $10^{-4}$. From these results, we conclude that the PINN can predict the positon solution of coupled NLSE which can further be confirmed from the various time snapshots between exact and predicted results given in Figs. \ref{8}(g)-(l).     

%
%
\begin{figure}[ht!]\centering\includegraphics[height=10cm,width=1\linewidth]{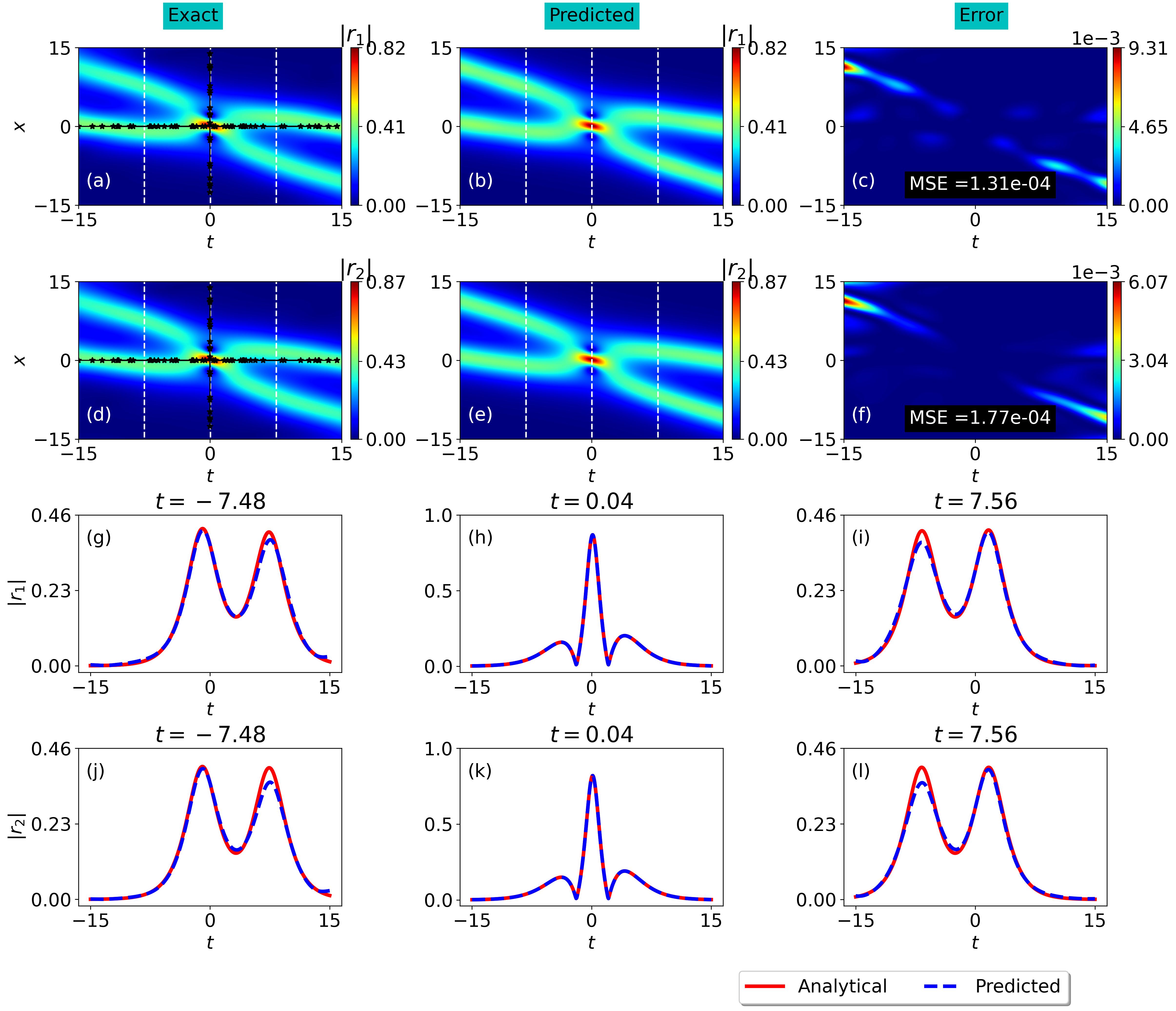}\caption{Predicted and exact solution of second order positon of coupled Hirota equation using PINN approach. (a),(d) Contour profile of analytical second order positon solution of two components $r_1$ and $r_2$ with parameter value $\lambda = 0.1+ 0.3 i$, $ s_1 =1.01 ,s_2 = 0.9, s_3 = 0.85$ and $\mu=0.1$. (b),(e) The predicted second order positon solution by PINN. (c),(f) Mean squared error plot between the analytical and predicted solution. (g)-(l) are the various time snapshots of predicted and exact solution respectively. The exact region of snapshots is denoted as white dotted lines in (a), (b), (d) and (e). The black line in (a) and (d) represents the data points taken at initial and new conditions, respectively.}\label{9}\end{figure}
\subsection{Data driven positon solution of coupled Hirota equation}
Finally, we explore the positon solution of the coupled Hirota equation through PINN approach. Tasgal and Potasek proposed the CH equations, incorporating higher-order effects such as third dispersion, self-steepening, and inelastic Raman scattering terms to depict a non-relativistic boson field \cite{TP}. The underlying equation plays a significant role in optics \cite{bindu}. The study of positon solution of Eq.\eqref{che} through PINN algorithm is yet to be investigated in the literature. We consider the CH equations  in the following form
\begin{equation}
\begin{aligned}
			i r_{1t} + r_{1xx} + 2 (|r_1|^2 + |r_2|^2)r_1 + i \mu (r_{1xxx} +(6 |r_1|^2 + 3|r_2|^2)r_{1x} + 3 r_1 r_2^* r_{2x})=0, \\
			i r_{2t} +r_{2xx} + 2 (|r_1|^2 + |r_2|^2)r_2 + i \mu (r_{2xxx} +(6 |r_2|^2 + 3|r_1|^2)r_{2x} + 3 r_2 r_1^* r_{1x})=0,
		\end{aligned}\label{che}
\end{equation} 
where $r_1$ and $r_2$ are the complex wave envelope functions with $\mu$ as the arbitrary nonlinear parameter. Eq.\eqref{che} can be converted to the coupled NLSE by letting $\mu=0$. Interaction among rogue waves and the other nonlinear waves such as dark-bright solitons and breathers of Eq.\eqref{che} are reported in \cite{DT-che}. In our earlier studies, we have constructed the positon and breather positon solutions of Eq.\eqref{che} through gDT method in \cite{nld-paper}. For the present analysis, we consider only the second order positon solution of Eq.\eqref{che}. To begin, we split the complex wave functions ($r_{1,2} = u_{1,2} + i v_{1,2}$) into the following equations  
\begin{eqnarray}
g_{u_{1}}(x,t) &:=& 2  u_1^3 + 2  u_1  u_2^2 + 2  u_1  v_1^2 + 2  u_1  v_2^2 - v_{1t} - 3 \mu u_2 v_1 u_{2x} + 3\mu u_1 v_2 u_{2x}\notag \\&&  - 6\mu u_1^2 v_{1x} - 3\mu u_2^2 v_{1x} - 6\mu v_1^2 v_{1x}  + u_{1xx}  - 3\mu v_2^2 v_{1x}\notag \\&&- 3\mu u_1 u_2 v_{2x}- 3\mu v_1 v_2 v_{2x}  - \mu v_{1xxx}\notag   \\
g_{v_{1}}(x,t) &:=& 2  u_1^2  v_1 + 2  u_2^2  v_1 + 2  v_1^3 + 2  v_1  v_2^2 + u_{2t} + 6\mu u_1^2 u_{1x}+ 3\mu u_2^2 u_{1x} \notag \\&&  + 6\mu v_1^2 u_{2x}  + 3\mu v_2^2 u_{1x} + 3\mu u_1 u_2 u_{2x} + v_{1xx}  + 3\mu v_1 v_2 u_{2x} \notag \\&& - 3\mu u_2 v_1 v_{2x} + 3\mu u_1 v_2 v_{2x} + \mu u_{1xxx}\notag \\
g_{u_{2}}(x,t) &:=& 2  u_2^3 + 2  u_2  u_1^2 + 2  u_2  v_2^2 + 2  u_2  v_1^2  - 3\mu u_1 v_2 u_{1x} + 3\mu u_2 v_1 u_{1x}\notag  \\&&  
          - 6\mu u_2^2 v_{2x} - 3\mu u_1^2 v_{2x} - 6\mu v_2^2 v_{2x}   - 3\mu v_1^2 v_{2x}\notag \\&& - 3\mu u_2 u_1 v_{1x} - 3\mu v_2 v_1 v_{1x}- v_{2t}+ u_{2xx} - \mu v_{2xxx}\notag 
 \\
g_{v_{2}}(x,t) &:=& 2  u_2^2  v_2 + 2  u_1^2  v_2 + 2  v_2^3 + 2  v_2  v_1^2 + 6\mu u_2^2 u_{2x} + 3\mu u_1^2 u_{2x}\notag  \\&& + 6\mu v_2^2 u_{2x}  + 3\mu v_1^2 u_{2x} + 3\mu u_1 u_2 u_{1x}+ 3\mu v_1 v_2 u_{1x}\notag 
         \\&&  - 3\mu u_1 v_2 v_{1x} + 3\mu u_2 v_1 v_{1x}+ u_{2t} + v_{2xx} + \mu u_{2xxx}.
\end{eqnarray}
 \par The initial and boundary data points are extracted from the second order positon solution of Eq. \eqref{che}. We adapt the same preliminary setup of PINN which we considered for the other equations. For the present case, we consider $x, t \in [-15, 15]$ as the spatial and time range and we run the model for $50000$ iterations to optimize the loss function. The second order positon solutions of both the components are predicted through PINN algorithm. The expected results of Eq.\eqref{che} are produced in Fig. \ref{9}. Subsequently, the exact and predicted solutions of $r_1$ and $r_2$ are shown in Figs. \ref{9} (a), (d) and  \ref{9} (b), (e), respectively. The error plot shows that the intensity has been reduced significantly, see Figs. \ref{9} (c), (f) which has the mean squared error value in the order of $10^{-4}$. The relationship between the predicted and actual positon solutions at different time instances is graphically represented in  Figs.\ref{9} (g)-(l) affirming the precision of PINN. 
\subsection{Data driven positon solution for the coupled generalized NLSE}

\begin{figure}[ht!]
\centering
\includegraphics[height=10cm,width=1\linewidth]{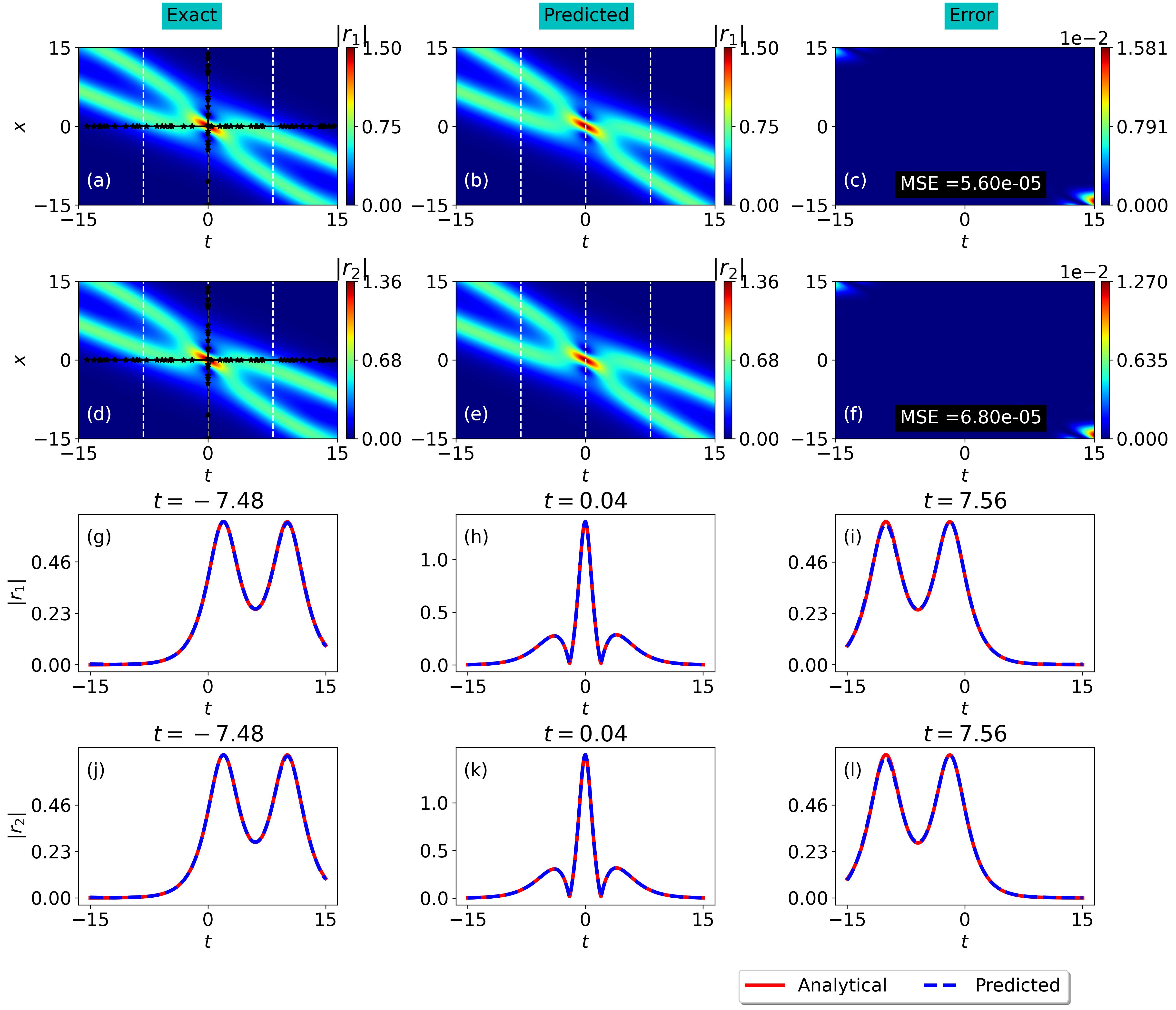}
\caption{Predicted and exact solution of second order positon of coupled generalized NLSE using PINN approach. (a),(d) Contour profile of analytical second order positon solution of two components $r_1$ and $r_2$ with parameter values $\lambda = 0.2+ 0.3 i$, $s_1 =1.2 ,s_2 = 0.9, s_3 = 0.85, k=0.2+0.4i, d=0.5, c=0.7 $ (b),(e) The predicted second order positon solution by PINN. (c),(f) Mean squared error plot between analytical and predicted solution. (g)-(l) are the various time snapshots of predicted and exact solution respectively. The exact region of snapshots is denoted as white dotted lines in (a), (b), (d) and (e). The black line in (a) and (d) represents the data points taken at initial and new conditions, respectively.}
\label{10}
\end{figure}

In this section, we investigate a system governed by the following coupled generalized NLSE \cite{gcnls}, 
\begin{equation}
\begin{aligned}
			i q_{1t} + q_{1xx} + 2 ( d |q_1|^2 + c |q_2|^2 k q_1 q_2^* + k^* q_2 q_1^* )r_1 =0, \\
			i q_{2t} + q_{2xx} + 2 ( d |q_1|^2 + c |q_2|^2 k q_1 q_2^* + k^* q_2 q_1^* )r_2 =0.
		\end{aligned}\label{gcnls0}
\end{equation} 
The wave functions are denoted by $r_1$ and $r_2$, while $t$ and $x$ are time and space variables. The constants $c$ and $d$ represent how the intensity of one wave can affect its own speed (self-phase modulation) and the speed of the other wave (cross-phase modulation). Additionally, the complex constant $k$ captures the phenomenon of four-wave mixing, where interacting nonlinear waves generate new wave types. The asterisk $*$ denotes the complex conjugate. It's important to note that under specific conditions $d = c$ and $k = 0$, the governing equation simplifies to the Manakov system. Analyzing the influence of the four-wave mixing parameter (represented by $d$) in coupled NLSEs has faced limitations. However, Agalarov et al.\cite{trans} recently introduced the transformation $q_1 = r_1 - k^{*} r_2, q_2 = d  r_2$ in Eq.\eqref{gcnls} to overcome the limitations. Then the Eq.\eqref{gcnls} becomes
\begin{equation}
\begin{aligned}
i r_{1t} + r_{1xx} + 2 d (  |r_1|^2 + \sigma |r_2|^2  )r_1 = 0, \\
i r_{2t} + r_{2xx} + 2 d ( |r_1|^2 + \sigma |r_2|^2 )r_2 = 0.
		\end{aligned}\label{gcnls}
\end{equation}
This transformation simplifies the coupled generalized NLSE \eqref{gcnls0} into the system presented in Eq~\eqref{gcnls}. Here, a new parameter $\sigma = d c - |k^2|$ emerges, which combines the original constants $d$, $c$, and $k$. Notably, when $d$ and $\sigma$ are both set to $1$, Eq.~\eqref{gcnls} reduces to the well-known Manakov model. Now, we can recognize that the four-wave mixing effect is implicitly embedded within Eq.~\eqref{gcnls}. As explored in \cite{gcnls}, this phenomenon plays a significant role in the complex interactions between various nonlinear waves, including solitons, breathers, and rogue waves. DT for this Eq.\eqref{gcnls} already exists in \cite{dtgcnls}, here, we employ generalized DT method and constructed the second order positon solution for the governing equation. The detailed calculations will be published in future. In this work, we focus on predicting these positon solution in PINN approach. First we split the complex wave functions into real and imaginary parts as $r_{1,2} = (u_{1,2} + i v_{1,2})$ and rewrite the Eq.\eqref{gcnls} in the following form 
\begin{eqnarray}
g_{u_{1}}(x,t) &:=& 2d u_1^3 + 4 k_r u_1^2 u_2 + 2 c u_1 u_2^2 - 4 k_i u_1 u_2 v_1 + 2d u_1 v_1^2 + 4 k_i u_1^2 v_1 \notag   \\&&+ 4k_r u_1 v_1 v_2 + 2c u_1 v_2^2 - v_{1t} + u_{1xx}
\notag   \\
g_{v_{1}}(x,t) &:=&2 d u_1^2 v_1 + 4k_r u_1 u_2v_1 + 2c u_2^2 v_1 - 4k_i u_2 v_1^2 + 2d v_1^3 + 4k_i u_1 v_1 v_2\notag   \\&& + 4 k_r v_1^2 v_2 + 2c v_1 v_2^2 + u_{1t} + v_{1xx}
 \notag \\
g_{u_{2}}(x,t) &:=& 2d u_1^2 u_2 + 4k_r u_1 u_2^2 + 2c u_2^3 - 4k_i u_2^2 v_1 + 2d u_2 v_1^2 + 4 k_i u_1 u_2 v_2 \notag   \\&&+ 4 k_r u_2v_1 v_2 + 2 c u_2 v_2^2 - v_{2t} + u_{1xx}\notag \\
g_{v_{2}}(x,t) &:=& 2du_1^2v_2 + 4k_r u_1 u_2 v_2 + 2c u_2^2 v_2 - 4k_i u_2 v_1 v_2 + 2d v_1^2 v_2 + 4k_i u_1 v_2^2 \notag   \\&&+ 4k_r v_1 v_2^2 + 2cv_2^3 + u_{2t} + v_{2xx}\notag   \\
\end{eqnarray}
Our investigation leverages the established PINN architecture with a 4-neuron output layer to handle the two-component system described by the coupled generalized NLSE \eqref{gcnls}. The spatial and temporal domain spans from $-15$ to $15$ for both position ($x$) and time ($t$). We employ the Adam optimizer for $30,000$ iterations during the training process. Using this optimized PINN, we aim to predict the second-order positon solution for the coupled NLSE. Figure \ref{10} presents the predicted results alongside the exact solutions for the two components. Figures \ref{10} (a), (b), (d), and (e) illustrate these comparisons. The mean squared error between the predicted and exact solutions is on the order of $10^{-5}$, as shown in Figures \ref{10}(c) and (f). These results demonstrate the capability of PINNs to predict positon solutions for coupled generalized NLSEs. This is further supported by the close agreement between the exact and predicted solutions across various time snapshots depicted in Figures \ref{10}(g) to (l).

%
\section{Conclusion}     
In this work, we have considered \textit{Physics-Informed Neural Networks} to investigate data-driven positon solutions within a family of higher-order nonlinear Schr{\"o}dinger Equations. We have demonstrated that with minimal information (data points from the solution at $t=0$ and $x=0$) one can predict the propagation of second-order positons in NLSEs with cubic, quartic, quintic nonlinearities. We have also extended the methodology for two-component systems, namely (i) coupled NLSE and (ii) coupled Hirota equations. In addition to the above studies, with the help of GDT method we have derived the second-order positon solutions for both the sixth-order NLSE and the coupled generalized NLSE. These two solutions are new to the literature. Then, we leveraged PINNs to predict the behaviour of these two solutions. Predicted outcomes closely align with the exact solutions obtained through the analytical methods. We have also compared the performance of our modified PINN algorithm with the traditional PINN approach for predicting positon solutions of NLSEs.
Our findings reveal that, for basic NLSE, both approaches achieve similar MSE values, suggesting comparable performance for lower-order equations. For higher-order NLSEs, the considered modified PINN algorithm significantly outperforms the traditional method, requiring less iterations to achieve high accuracy. The traditional PINN algorithm requires more number of iterations to achieve the same level of accuracy. Some of the NLSEs exhibit similar MSE values for both methods, so we did not present the visual representation of PINN results with Dirichlet boundary condition, as it would not provide significant insights. Hence we present the comparison of MSE values obtained using both PINN methods for various NLSEs in Table.~1. 
\begin{table}
    \centering
    \begin{tabular}{|c|c|c|c|}
        \hline  PDEs&   MSE (for Dirichlet BC) & MSE (for $x=0$ condition)  \\ \hline \rule{0pt}{3ex}
        NLSE &  $3.10 \times 10^{-7}$ & $6.20 \times 10^{-7}$  \\
         Hirota equation& $1.70 \times 10^{-5}$ & $1.60 \times10^{-5}$  \\
        Fourth order NLSE & $1.00 \times 10^{-4}$ & $2.00 \times10^{-6}$  \\
        Extended NLSE &  $2.00 \times10^{-2}$ & $2.00 \times 10^{-4}$  \\
        Fifth order NLSE &  $8.00 \times 10^{-4}$  & $8.00 \times 10^{-4}$ \\
        Sixth order NLSE &  $8.00 \times10^{-3}$ & $4.00 \times 10^{-4}$  \\\rule{0pt}{3ex}
        Coupled NLSE &  $8.50 \times10^{-4},~~ 1.29 \times10^{-3}$ & $8.50 \times 10^{-5},~~ 1.20 \times10^{-4}$\\
      Coupled Hirota  &   $1.31 \times 10^{-3}, ~~1.77 \times10^{-3}$ & $1.31 \times 10^{-4},~~ 1.77 \times 10^{-4}$  \\
      Coupled generalized NLSE  & $5.60 \times10^{-5},~~ 6.80 \times 10^{-5}$ & $3.80 \times10^{-5},~~ 4.90 \times 10^{-5}$  \\\hline
    \end{tabular}
\caption{Calculated mean squared error values from the results of PINN with Dirichlet boundary condition and $x=0$ condition for the family of NLSEs}
\end{table}
As shown in the table, the traditional PINN method requires further improvement for consistent performance across various PDEs and this data confirms the improved efficiency and accuracy of our modified approach for higher-order PDES. These results also demonstrate the effectiveness of our modifications in addressing the limitations identified in the original PINN algorithm, particularly for handling the complexities of higher-order nonlinear PDEs. Visual comparisons of predicted and exact solutions at various time points further confirm the method's precision. This work achieves a significant breakthrough by demonstrating the PINN's capability to solve PDEs with nonlinearities of up to sixth order, surpassing limitations cited in prior literature \cite{tonls}. Our results hold promise for future applications in the numerical and PINN communities. Further exploration will focus on using deep learning to investigate breather positon and hybrid solutions in nonlinear integrable systems.

\appendix
\section{}
The exact expression for the second order positon solution of NLS equation is given by \cite{w1}
\begin{eqnarray}
	r = \dfrac{A_1}{D_1},\label{nls1}
\end{eqnarray}
where 
\begin{eqnarray}
A_1 &=& 8 b e^{2 (b-i a) (2 a t + 2 i b t + x)} (1 + 8 a b t + 8 i b^2 t + 
2 b x + e^{4 b (4 a t + x)} (1 - 8 a b t \notag\\&&
+ 8 i b^2 t - 2 b x)), \notag\\
 D_1& =& 1 + e^{8 b (4 a t + x)} + 
 2 e^{4 b (4 a t + x)} (1 + 128 a^2 b^2 t^2 + 128 b^4 t^2 + 
 64 a b^2 t x + 8 b^2 x^2). \notag
\end{eqnarray}
\par The second order positon solution of third order NLS equation \eqref{hirota} takes the following form \cite{w1}
\begin{equation}
r= \dfrac{A_2}{D_2},\label{hirota1}
\end{equation}
with
\begin{eqnarray}
A_2 &=& 8 b e^{(4 (a -i b)^2 (-i + 2i a + 2 b) t + 2i a x - 
2 b (8 a t + 8 b^2 t + x)) }(e^{(
4 b (4 a t + 4 b^2 t + x))} (1 + 24 a^2 b t\notag\\&& 
+ 8i b^2 t - 24 b^3 t - 2 b x) + 
e^{(48 a^2 b t)} (1 - 24 a^2 b t + 8 a (1 - 6i b) b t\notag\\&&  + 
8 a (-1 - 6i b) b t + 8i b^2 t + 
24 b^3 t + 2 b x)), \notag \\
B_2 & =& e^{(4 (i a + b) x)} + e^{(-32 a b t + 96 a^2 b t - 32 b^3 t + 4i a x - 4 b x)} + 2 e^{(-16 a b t + 48 a^2 b t - 16 b^3 t + 
4i a x)} \notag\\&& 
\times (1 - 768 a^3 b^2 t^2 + 1152 a^4 b^2 t^2 + 1152 b^6 t^2  + 
64 a^2 b^2 t ((2 + 36 b^2) t - 3 x) \notag\\&& + 8 b^2 x^2+ 
64 a b^2 t (-12 b^2 t + x) + 64 b^4 t (2 t + 3 x)). \notag
\end{eqnarray}
\par The following expression represents the second order positon solution of the fourth order NLS equation \eqref{lpd} \cite{w1}
\begin{equation}
	r= \dfrac{ A_3}{D_3},\label{lpd1}
\end{equation}
where
\begin{eqnarray}
A_3 & = &  8 b e^{2 (-2i (a -i b)^2 + 8i (a -i b)^4) t - 
4i a x} (e^{
2 (i a + b) x} (1 + 64 a^3 b t + 8i b^2 t - 192i a^2 b^2 t \notag\\&& + 
64i b^4 t - 8 a (b + 24 b^3) t - 2 b x) + 
e^{(2 a (8 (-1 + 8 a^2) b t - 64 b^3 t +i x) - 
2 b x)}  \notag\\&& \times (1 - 64 a^3 b t+ 8i b^2 t - 192i a^2 b^2 t + 
64i b^4 t + 8 a (b + 24 b^3) t + 2 b x)), \notag \\
D_3 & = & e^{4 b x} + e^{(-4 b (-64 a^3 t + 8 a (t + 8 b^2 t) + x))} + 
2 e^{(16 a b (-1 + 8 a^2 - 8 b^2) t)} (1 + 8192 a^6 b^2 t^2 \notag\\&& + 
128 b^4 t^2 + 2048 b^6 t^2 + 8192 b^8 t^2 + 
2048 a^4 b^2 (-1 + 12 b^2) t^2  \notag\\&&+ 128 a^2 b^2 (1 + 192 b^4) t^2- 
512 a^3 b^2 t x + 64 a b^2 (1 + 24 b^2) t x + 8 b^2 x^2). \notag
\end{eqnarray}
\par The second order positon solution of extended NLS equation \eqref{enls} is given by \cite{w2}
\begin{equation}
	r= \dfrac{ A_4}{D_4},\label{enls1}
\end{equation}
where
\begin{eqnarray}
A_4 & = & 8 b e^{-8 i (a - i b)^2 (1/2 - a - 2 (a - i b)^2 + i b) t + 2 i a x - 2 b (8 b^2 t + x)} (e^{
4 b (4 b^2 t + x)} (1 + 
8 i b (i a + b)\notag\\&& \times (1 - 8 a^2 + a (-3 + 16 i b) + 
b (3 i + 8 b)) t - 2 b x) + 
e^{(16 a b (-1 + a (3 + 8 a) - 8 b^2) t)} \notag\\&& \times (1 + 
2 b (-4 (a + i b) (-1 + 8 a^2 + 
a (3 + 16 i b) + (3 i - 8 b) b) t + x))),\notag\\
D_4 & = &e^{4 (i a + b) x} + 
e^{-32 b (b^2 + a (1 - a (3 + 8 a) + 8 b^2)) t + 4 i (a + i b) x} \notag\\&&+ 
e^{-16 b (b^2 + a (1 - a (3 + 8 a) + 8 b^2)) t + 
4 i a x}  (2 + 
16 b^2 (16 (a^2 + 
b^2) ((-1 + a (3 + 8 a))^2 \notag\\&&+ (25 + 16 a (3 + 8 a)) b^2   + 64 b^4) t^2+ 8 (3 b^2 + a (1 - a (3 + 8 a)\notag\\&& + 24 b^2)) t x + x^2)).\notag 
\end{eqnarray}
\par The second order positon solution of fifth order NLS equation \eqref{5nls} takes the following form \cite{fnls}

\begin{equation}
r = \dfrac{A_5}{D_5},\label{fnls1}
\end{equation}
where
\begin{eqnarray}
A_5 &=& -4 b e^{-4 i (2 a^2 t - 2 b^2 t + a x + 
16 a (a^4 - 10 a^2 b^2 + 5 b^4) t \delta)} (-e^{
2 i (a + i b) (2 a t + 2 i b t + x) + 
32 i (a + i b)^5 t \delta}\notag\\
&& \times  (1 + 
2 b (4 a t + 4 i b t + x + 80 (a + i b)^4 t \delta)) + 
e^{2 (i a + b) (2 a t - 2 i b t + x + 
16 (i a + b)^4 t \delta)}\notag\\&& \times (-1 + 
2 b (4 a t - 4 i b t + x + 80 (i a + b)^4 t \delta))), \notag \\
D_5 & = & (1 + 8 b^2 (16 a^2 t^2 + 16 b^2 t^2 + 8 a t x + x^2 + 
160 t (4 a (a^2 - 3 b^2) (a^2 + b^2) t   \notag\\&&+ (a^4 - 6 a^2 b^2 + 
b^4) x) \delta + 6400 (a^2 + b^2)^4 t^2 \delta^2) \notag\\&&+ 
\cosh[4 b (x + 4 t (a + 4 (5 a^4 - 10 a^2 b^2 + b^4) \delta))]). \notag
\end{eqnarray}

\par The second order positon solution of Sixth-order NLS equation \eqref{6nls} takes the following form 

\begin{equation}
r = \dfrac{A_6}{D_6},\label{6nls1}
\end{equation}
where
\begin{eqnarray}
A_6 &=& 8 b e^{-4 i (a x + 
    2 (a^2 - b^2) t (1 + 16 a^4 \xi - 224 a^2 b^2 \xi + 
       16 b^4 \xi))} (i e^{
    2 i (a - i b) (2 (a - i b) t + x + 32 (a - i b)^5 t \xi)}  \notag \\&& \times (-i + 
      2 i b x + 8 i (a - i b) b t (1 + 48 (a - i b)^4 \xi)) + 
   e^{2 i (a + i b) (2 (a + i b) t + x + 
       32 (a + i b)^5 t \xi)} \notag \\&& \times (1 + 2 b x + 
      8 (a + i b) b t (1 + 48 (a + i b)^4 \xi))),\notag \\
D_6 & = &2 + e^{-4 b (x + 192 a^5 t \xi - 640 a^3 b^2 t \xi+ 
    4 a (t + 48 b^4 t \xi))} + e^{
 4 b (x + 192 a^5 t \xi - 640 a^3 b^2 t \xi + 
    4 a (t + 48 b^4 t \xi))} \notag \\&&+ 16 b^2 x^2 + 
 128 a b^2 t x (1 + 48 a^4 \xi - 480 a^2 b^2 \xi + 
    240 b^4 \xi) + 
 256 (a - i b)\notag \\&& \times (a + i b) b^2 t^2 (1 + 48 (a - i b)^4 \xi + 
    48 (a + i b)^4 \xi (1 + 48 (a - i b)^4 \xi)).  \notag
\end{eqnarray}

\par The second order positon solution of coupled NLS equation \eqref{cnls} is given by \cite{cnls}
\begin{equation}
r_1= \dfrac{ s_2 A_7}{D_7}, \quad r_2= \dfrac{ s_3 A_7}{D_7} \label{cnls1}
\end{equation}
where 
\begin{eqnarray}
A_7 & = & 8 b e^{(4 i a^2 t + 8 a b t + 4 i b^2 t + 2 i a x + 
2 b x)} s_1 (e^{(4 b (4 a t + x))}
s_1^2 (-1 + 2 b (4 a t - 4 i b t + x)) \notag \\&& - (s_2^2 + s_3^2) (1 + 
2 b (4 a t + 4 i b t + x))),\notag\\
D_7 & = & e^{(8 i a^2 t + 32 a b t + 4 i a x + 8 b x)} s_1^4 + 
e^{(4 i a (2 a t + x))} (s_2^2 + s_3^2)^2 + 
2 e^{16 a b t + 4 b x + 4 i a (2 a t + x)} \notag \\&& \times
s_1^2 (s_2^2 + s_3^2) (1 + 8 b^2 (16 (a^2 + b^2) t^2 + 8 a t x + x^2)).
\end{eqnarray}
\par The coupled Hirota equation \eqref{che} constitutes the following second order positon solution \cite{nld-paper}
\begin{equation}
	r_1= \dfrac{s_2 A_8}{D_8}, \quad r_2= \dfrac{s_3 A_8}{D_8}\label{che1}
\end{equation}
where 
\begin{eqnarray}
	A_8 & = & -8 b s_1^3 (e^{
	2 i (a - i b) (x + 
	2 (a -i b) t (1 + 2 a \mu- 2i b \mu))}
	s_1^2 (1 - 2 b x - 
	8 (a -i b) b t\notag\\&& \times (1 + 3 a \mu - 3i b\mu))  + 
	e^{(2i (a +i b) (x + 
	2 (a +i b) t (1 + 2 a \mu+ 2i b\mu)))} (s_2^2 + 
	s_3^2) \notag\\&& \times(1 + 2 b x + 
	8 (a +i b) b t (1 + 3 a \mu+ 3i b\mu))), \notag\\
	D_8 & = & e^{4i (a -i b) (x + 
	2 (a -i b) t (1 + 2 a \mu- 2i b\mu))} s_1^6 + 
	e^{4i (a +i b) (x + 
	2 (a +i b) t (1 + 2 a \mu+ 2i b\mu))}\notag\\&& \times
	s_1^2 (s_2^2 + s_3^2)^2  - 
	2 e^{4i (2 a^2 t - 2 b^2 t + 4 a^3 t \mu+ 
	a (x - 12 b^2 t\mu))}
	s_1^4 (s_2^2 + s_3^2) (-1 - 8 b^2 x^2 \notag\\&&- 
	128 (a -i b) (a +i b) b^2 t^2(1 + 3 a \mu- 
	3i b\mu) (1 + 3 a \mu+ 3i b\mu) \notag\\&& + 
	64 b^2 t x (-a - 3 a^2 \mu+ 3 b^2\mu)).
\end{eqnarray}
 \par The second order positon solution of coupled generalized NLSE \eqref{gcnls} is given by
\begin{equation}
r_1= \dfrac{ s_2 A_9}{\sqrt{d}  D_9}, \quad r_2= \dfrac{ s_3 A_9}{\sqrt{d(dc - k k^*)}D_9} \label{gcnls1}
\end{equation}
where 
\begin{eqnarray}
	A_9 & =& 8 i b s_1 (-i e^{(2 (i a + b) (2 a t - 2 i b t + x))}
     s_1^2 (-1 + 8 a b t - 8 i b^2 t + 2 b x) \notag\\&&+ 
   i e^{(2 i (a + i b) (2 a t + 2 i b t + x))} (s_2^2 + s_3^2) (1 + 
      8 a b t + 8 i b^2 t + 2 b x)) , \notag\\
	D_9 & = & (e^{4 (i a + b) (2 a t - 2 i b t + x)} s_1^4 + 
   e^{4 i (a + i b) (2 a t + 2 i b t + x) }(s_2^2 + s_3^2)^2 + 
   2 e^{4 i (2 a^2 t - 2 b^2 t + a x)}\notag\\&&
     s_1^2 (s_2^2 + s_3^2) (1 + 128 a^2 b^2 t^2 + 128 b^4 t^2 + 
      64 a b^2 t x + 8 b^2 x^2)) .\notag\\
\end{eqnarray} 
\section*{Acknowledgement} KT wishes to thank DST-SERB, Government of India, for providing the student internship under Grant No. CRG/2021/002428. NVP expresses gratitude to the Department of Science and Technology, India, for financial support through the Women Scientist Scheme-A under the Grant No. DST/WOS-A/PM-95/2020. SM acknowledges the fellowship provided by MoE-RUSA 2.0 Physical Sciences, Government of India, for conducting this research. The efforts of MS were backed by DST-SERB, Government of India, under Grant No. CRG/2021/002428.


\bibliographystyle{elsarticle-num}

\end{document}